\documentclass[aps,prd,amsmath,preprint]{revtex4}

\usepackage{amsmath}
\usepackage{bm}
\usepackage[dvips]{color,graphicx}
\usepackage{epsfig}
\usepackage{dcolumn}

\newcommand{\be}{\begin{equation}}
\newcommand{\ee}{\end{equation}}

\newcommand{\gtsim}{\lower .7ex\hbox{$\;\stackrel{\textstyle >}{\sim}\;$}}
\newcommand{\lapproxeq}{\lower .7ex\hbox{$\;\stackrel{\textstyle <}{\sim}\;$}}
\newcommand{\fslash}[1]{\mbox{$\!\not\!#1$}}
\newcommand{\ubar}{\bar{u}}
\newcommand{\sw}{\sin^2\theta_W}
\newcommand{\cw}{\cos^2\theta_W}
\newcommand{\eps}{\varepsilon}

\preprint{JLAB-THY-09-956}

\begin{document}

\title{Detailed Analysis of Two-Boson Exchange	\\
	in Parity-Violating $e$--$p$ Scattering}

\author{J. A. Tjon}
\affiliation{Physics Department, University of Utrecht,
        The Netherlands}

\author{P. G. Blunden}
\affiliation{Department of Physics and Astronomy,
        University of Manitoba,
        Winnipeg, MB, Canada R3T 2N2}

\author{W. Melnitchouk}
\affiliation{Jefferson Lab, 12000 Jefferson Avenue,
        Newport News, Virginia 23606, USA}

\begin{abstract}
We present a comprehensive study of two-boson exchange (TBE)
corrections in parity-violating electron--proton elastic scattering.
Within a hadronic framework, we compute contributions from box
(and crossed box) diagrams in which the intermediate states are
described by nucleons and $\Delta$ baryons.
The $\Delta$ contribution is found to be much smaller than the
nucleon one at backward angles (small $\eps$), but becomes dominant
in the forward scattering limit ($\eps \to 1$), where the nucleon
contribution vanishes.
The dependence of the corrections on the input hadronic form factors
is small for $Q^2 \lesssim 1$~GeV$^2$, but becomes significant at
larger $Q^2$.
We compute the nucleon and $\Delta$ TBE corrections relevant for
recent and planned parity-violating experiments, with the total
corrections ranging from $-1\%$ for forward angles to $1-2\%$ at
backward kinematics.
\end{abstract}

\maketitle

\section{Introduction}

Parity-violating electron--proton elastic scattering has become a
standard tool with which to probe the strangeness content of the proton.
Recent high-precision experiments at Jefferson Lab
\cite{HAPPEX04,G0,HAPPEX07,G0back} and elsewhere 
\cite{SAMPLE97,SAMPLErev,PVA404,A4back} have provided important
constraints on the strange electric and magnetic form factors
\cite{Young,MRM07}.
Further improvements in the precision are expected to allow the
measurement of the proton's weak charge,
$Q_w = 1-4\sw$, where $\theta_W$ is the weak mixing angle,
to unprecedented accuracy \cite{QWEAK,YoungSM}.

With the increasing precision comes the need to understand backgrounds
to greater accuracy than was called for in previous generations of
experiments.
In particular, higher-order radiative effects have received renewed
attention recently, most notably those associated with the exchange of
two bosons (photons or $Z$-bosons)
\cite{Marciano,MS,MRM,AC,ABB,Yang,TM,YangD}.
For point-like particles, the relevant loop diagrams are straightforward
to compute and are included in the standard radiative corrections.
However, incorporating the finite size of the nucleon leads to additional
contributions, and can introduce further uncertainty in the calculations.

In electromagnetic elastic scattering, despite being ${\cal O}(\alpha)$
suppressed, two-photon exchange (TPE) was found to play an important role
in resolving a large part of the discrepancy between the electric to
magnetic proton form factor ratio measurements using the Rosenbluth
and polarization transfer methods (see Ref.~\cite{BMT05} and references
therein).
One needs to carefully consider, therefore, to what extent the hadronic
structure effects in two-boson exchange (TBE) may affect the analysis
of parity-violating electron scattering.
This is especially critical given that the extracted strange form factors
appear to be rather small \cite{Young}, as is the proton's weak charge
$Q_w$, which could further enhance the relative importance of TBE effects.

In their seminal early work on electroweak radiative effects, Marciano
\& Sirlin \cite{Marciano} computed the interference between the
one-photon exchange and $\gamma$--$Z$ exchange amplitudes (which we
denote by ``$\gamma(Z\gamma)$'') at zero four-momentum transfer squared
$Q^2$, both at the quark level and at the nucleon level using dipole
form factors.
The corresponding contribution from the interference between the
single $Z$-boson and two-photon exchange amplitudes (denoted by
``$Z(\gamma\gamma)$'') vanishes at $Q^2=0$,
but was computed within a generalized parton distribution formalism
\cite{AC} at a scale $Q^2 \sim$ several GeV$^2$.

More recently, the TBE corrections were computed at nonzero $Q^2$
in a hadronic basis, including nucleon \cite{Yang,TM} and $\Delta$
\cite{YangD} intermediate states, with the structure dependence
incorporated through hadronic form factors.
For the nucleon intermediate states the model dependence was studied
in Ref.~\cite{TM}, and the individual TBE corrections to the proton
and neutron terms in the parity-violating asymmetry computed.

In this paper we perform a detailed analysis of TBE including both
nucleon elastic and $\Delta$ intermediate states in the loop diagrams,
and carefully examine their model dependence.
We use the hadronic formalism developed in Ref.~\cite{BMT03}, which
allows a natural implementation of hadronic structure effects in
radiative corrections at low $Q^2$, where parity-violating electron
scattering experiments are typically performed.
For the $\Delta$ contribution we extend the two-photon exchange
calculation of Kondratyuk {\em et al.} \cite{KondD} to the weak sector,
and constrain the axial-vector form factors by data from neutrino
scattering.

In Sec.~\ref{sec:Born} we review the basic formalism of parity-violating
electron scattering and summarize the Born level amplitudes and cross
sections.
The two-boson exchange corrections are described in Sec.~\ref{sec:TBE},
where we outline the box diagram calculations with nucleon and $\Delta$
intermediate states.
Our main results are presented in Sec.~\ref{sec:res}.
We compute the corrections from TBE to the parity-violating asymmetry,
and discuss the consequences for the extraction of the proton's strange
form factors and weak axial charge.
Finally, we summarize our findings in Sec.~\ref{sec:conc} and identify
possible future developments of this work.

\section{Born Approximation}
\label{sec:Born}

For elastic scattering of an electron $e^-$ from a nucleon $N$ we
define the initial $e^-$ and $N$ momenta as $p_1$ and $p_2$,
and final $e^-$ and $N$ momenta as $p_3$ and $p_4$, respectively,
$e^-(p_1) + N(p_2) \to e^-(p_3) + N(p_4)$.
The four-momentum transferred from the electron to the nucleon
is given by $q = p_4-p_2 = p_1-p_3$, with $Q^2 \equiv -q^2 > 0$.
In the Born approximation, the amplitudes for the electromagnetic
and weak neutral currents are given by:
\begin{eqnarray}
{\cal M}_\gamma
&=& -{e^2 \over q^2}\
    j_\gamma^\mu\ J_{\gamma \mu}\ , \\
{\cal M}_Z
&=& -{g^2 \over (4 \cos\theta_W)^2}\
    {1 \over M_Z^2-q^2}\
    j_Z^\mu\ J_{Z \mu}\
\approx\ -{G_F \over 2\sqrt{2}}\ j_Z^\mu\ J_{Z \mu}\  ,
\end{eqnarray}
where $e$ is the electric charge, $g = e/\sin\theta_W$ is the
weak coupling constant, $M_Z$ is the $Z$ boson mass, and
$G_F = \pi\alpha/(\sqrt{2} M_Z^2 \sw\cw)$ is the Fermi constant,
with $\alpha = e^2/4\pi$ the fine structure constant.
At tree level the weak mixing angle is related to the weak boson
masses by $\sw = 1 - M_W^2/M_Z^2$, where $M_W$ is the $W$ boson mass
(in our numerical results below we use the renormalized value
$\sw = 0.2312$ \cite{PDG}).
The matrix elements of the electromagnetic and weak leptonic currents
are given by
\begin{eqnarray}
j_\gamma^\mu
&=& \ubar_e(p_3) \gamma^\mu u_e(p_1)\ , \\
j_Z^\mu
&=& \ubar_e(p_3) \left( g^e_V \gamma^\mu + g^e_A \gamma^\mu \gamma_5
         \right) u_e(p_1)\ ,
\end{eqnarray}
where the latter is given by a sum of vector and axial-vector terms.
We use the convention in which the vector and axial-vector couplings
of the electron to the $Z$ boson are given by
\begin{equation}
g^e_V\ =\ -(1-4 \sw)\ , \ \ \ \ \
g^e_A\ =\ +1\ .
\end{equation}
The matrix elements of the electromagnetic (weak) hadronic currents
can be written as
\begin{eqnarray}
J_{\gamma (Z)}^\mu
&=& \ubar_N(p_4)\ \Gamma_{\gamma (Z)}^\mu\ u_N(p_2)\ ,
\end{eqnarray}
where the current operators are parameterized by the electromagnetic
and weak form factors:
\begin{eqnarray}
\Gamma_\gamma^\mu
&=& \gamma^\mu\ F_1^{\gamma N}(Q^2)
 + {i \sigma^{\mu\nu} q_\nu \over 2 M} F_2^{\gamma N}(Q^2)\ , \\
\Gamma_Z^\mu
&=& \gamma^\mu\ F_1^{Z N}(Q^2)
 + \frac{i \sigma^{\mu\nu} q_\nu}{2 M}\ F_2^{Z N}(Q^2)
 + \gamma^\mu \gamma_5\ G_A^{Z N}(Q^2)\ ,
\label{eq:JZ}
\end{eqnarray}
with $M$ the nucleon mass.
Here $F_1$ and $F_2$ are the Dirac and Pauli form factors,
and $G_A$ the axial form factor of the nucleon ($N = p, n$), for
either the electromagnetic ($\gamma$) or weak ($Z$) current.
Usually one takes linear combinations of the Dirac and Pauli form
factors to define the Sachs electric and magnetic form factors as
\begin{eqnarray}
G_E(Q^2) &=& F_1(Q^2) - \tau F_2(Q^2)\ ,        \\
G_M(Q^2) &=& F_1(Q^2) + F_2(Q^2)\ ,
\end{eqnarray}
where $\tau = Q^2/4M^2$.

The differential cross section is given by the square of the sum of
the $\gamma$ and $Z$ Born amplitudes,
\be
{ d\sigma \over d\Omega }
= \left( { \alpha \over 4 M Q^2 } { E_3 \over E_1 } \right)^2\
  \left| {\cal M} \right|^2\ ,
\ee
where the squared amplitude can be written as
\be
\label{eq:Mamp}
\left| {\cal M} \right|^2
= \left| {\cal M}_\gamma + {\cal M}_Z \right|^2
= \left| {\cal M}_\gamma \right|^2
+ 2 \Re \left( {\cal M}^*_\gamma {\cal M}_Z \right)
+ \left| {\cal M}_Z \right|^2\ .
\ee
The purely weak contribution $|{\cal M}_Z|^2$ is small compared with
the other terms and can be neglected.
By polarizing the incident electron and measuring the difference
between right- and left-handed electrons scattering from unpolarized
protons, the parity-violating (PV) asymmetry can be defined in terms
of the differential cross sections as
\be
\label{eq:Apv}
A_{\rm PV} = {\sigma_R - \sigma_L \over \sigma_R + \sigma_L}\ ,
\ee
where $\sigma_{R(L)}$ is the cross section for a right- (left-)
hand polarized electron.
The purely electromagnetic contribution cancels in the numerator,
so that the asymmetry is sensitive to the parity-violating part of
$2\Re\left( {\cal M}^*_\gamma {\cal M}_Z \right)$, involving the
interference of ${\cal M}_\gamma$ with the product of vector and
axial-vector currents in ${\cal M}_Z$ (the vector-vector and
axial-axial parts of ${\cal M}_Z$ cancel in the asymmetry).
The denominator is dominated by the electromagnetic term,
$|{\cal M_\gamma}|^2$.

More explicitly, the PV asymmetry can be written in terms of the
electroweak form factors as
\begin{equation}
A_{\rm PV}
= -\left( {G_F Q^2 \over 4\sqrt{2}\pi\alpha} \right)
 { g_A^e \left( \eps G_E^{\gamma N} G_E^{Z N}
          + \tau G_M^{\gamma N} G_M^{Z N}
     \right)\
   +\ g_V^e\ \eps'\ G_M^{\gamma N} G_A^{Z N}
 \over
   \eps (G_E^{\gamma N})^2 + \tau (G_M^{\gamma N})^2
 }\ ,
\end{equation}
where $\eps$ and $\eps'$ are kinematical parameters,
\begin{eqnarray}
\eps^{-1} &=& 1 + 2 (1+\tau)\tan^2{\theta\over 2}\ , \\
\eps' &=& \sqrt{\tau (1+\tau)(1-\eps^2)}\ ,
\end{eqnarray}
with $\theta$ the electron scattering angle in the target rest frame.

For a proton target the weak electric (magnetic) vector form factor
$G_{E (M)}^{Z p}$ can be related by isospin symmetry to the
electromagnetic form factors of the proton and neutron by
\begin{equation}
\label{eq:GEMZ}
G_{E (M)}^{Z p}
= (1-4\sw) G_{E (M)}^{\gamma p} - G_{E (M)}^{\gamma n} - G_{E (M)}^s\ ,
\end{equation}
where $G_{E (M)}^s$ are the contributions from strange quarks.
The small factor $(1-4\sw)$ suppresses the overall contribution
from the proton electromagnetic form factors, thereby promoting the
neutron form factors to play a greater role.
The weak axial-vector form factor of the proton is given by
$G_A^{Z p} = -G_A^p + G_A^s$, where $G_A^s$ is the strange quark
contribution.

Measurement of the PV asymmetry $A_{\rm PV}$ as a function of the
scattering angle $\theta$ allows one to extract combinations of the
strange form factors, given knowledge of the proton and neutron
electromagnetic form factors.
Reliable extractions of the form factors require precise knowledge
of the radiative corrections to the PV scattering associated with
higher order electroweak processes.
This is especially critical given that the extracted strange form
factors appear to be rather small numerically.
In the next section we discuss a subset of the radiative corrections,
namely those arising from two-boson exchange.

\section{Two-Boson Exchange Corrections}
\label{sec:TBE}

Beyond the Born approximation, the PV asymmetry receives corrections
from higher order radiative effects, such as vertex corrections, wave
function renormalization, vacuum polarization, and inelastic
bremsstrahlung, which are well known and included in standard data
analyses.
Less well determined are radiative corrections arising from the
interference of Born and TBE diagrams,
both electromagnetic ($\gamma\gamma$) and electroweak ($\gamma Z$).
For purely electromagnetic scattering, the TPE corrections these have
been shown \cite{BMT05,BMT03} to display strong angular dependence,
which significantly affects extractions of the
$G_E^{\gamma p}/G_M^{\gamma p}$ ratio by Rosenbluth separation
\cite{AMT}.

There are several ways in which the PV asymmetry can be represented
in the presence of higher-order radiative corrections.
The approach pioneered by Marciano \& Sirlin \cite{MS} parameterizes
the electroweak radiative effects in terms of parameters $\rho$ and
$\kappa$, such that the weak charge of the proton in the presence of
higher order corrections becomes
\be
Q_w = 1-4\sw\ \to\ \rho (1-4\kappa\sw)\ .
\ee
In this case the asymmetry can be written as a sum of proton vector,
strange vector, and axial-vector contributions,
\begin{eqnarray}
A_{\rm PV}
= - \left( {G_F Q^2 \over 4\sqrt{2} \pi \alpha} \right)
    \left( A_V + A_s + A_A \right)\ ,
\label{eq:APV0}
\end{eqnarray}
where
\begin{subequations}
\label{eq:A}
\begin{align}
\label{eq:AV}
A_V
&= g^e_A\ \rho
\left[
  (1-4\kappa\sw)
- {1 \over \sigma_{\rm red}}
  \left( \eps G_E^{\gamma p} G_E^{\gamma n}
       + \tau G_M^{\gamma p} G_M^{\gamma n}
  \right)
\right],                \\
\label{eq:As}
A_s
&= - g^e_A\ \rho\
{1 \over \sigma_{\rm red}}
\left( \eps G_E^{\gamma p} G_E^s
     + \tau G_M^{\gamma p} G_M^s
\right)\ ,          \\
\label{eq:AA}
A_A
&= g^e_V\ \eps'\
{1 \over \sigma_{\rm red}}\
\widetilde{G}_A^{Z p} G_M^{\gamma p}\ ,
\end{align}
\end{subequations}
with
$\sigma_{\rm red} = \eps (G_E^{\gamma p})^2 + \tau (G_M^{\gamma p})^2$
the reduced unpolarized proton cross section.

An alternative parameterization is in terms of isoscalar and isovector
weak radiative corrections for the vector form factors, and a similar
set of corrections for the axial-vector form factors.
In this case the vector part of the PV asymmetry is written
\be
A_V = g^e_A
\left[ (1-4\sw) (1+R_V^p)
     - {1 \over \sigma_{\rm red}}
       \left( \eps G_E^{\gamma p} G_E^{\gamma n}
        + \tau G_M^{\gamma p} G_M^{\gamma n}
       \right) (1+R_V^n)
\right]\ ,
\ee
where the proton and neutron radiative corrections are given, to first
order in $\rho-1$ and $\kappa-1$, by
\begin{subequations}
\begin{eqnarray}
R_V^p &=& \rho - 1 - (\kappa-1)\ {4\sw \over 1-4\sw}\ , \\
R_V^n &=& \rho - 1\ .
\end{eqnarray}
\end{subequations}
The strange part of the asymmetry,
\be
A_s = - g^e_A
{1 \over \sigma_{\rm red}}
\left( \eps G_E^{\gamma p} G_E^s
     + \tau G_M^{\gamma p} G_M^s
\right) (1+R_V^{(0)})\ ,
\ee
receives an isoscalar radiative correction, given by
\begin{eqnarray}
R_V^{(0)} &=& \rho - 1\ .
\end{eqnarray}
For the axial asymmetry $A_A$, the form factor $\widetilde{G}_A^{Z p}$
implicitly contains higher order radiative corrections for the proton
axial current, as well as the hadronic anapole contributions
\cite{Young,Musolf}.
At tree level, and in the absence of the anapole term,
$\widetilde{G}_A^{Z p} \to G_A^{Z p}$.

In Refs.~\cite{Yang,TM,YangD} the contributions to $\rho$ and $\kappa$
from the interference of the Born and TBE (box and cross-box) diagrams
were computed, denoted by $\Delta\rho$ and $\Delta\kappa$, respectively.
The correction to the PV cross section arising from the the
$\gamma\gamma$ and $\gamma Z$ TBE contributions can be obtained from
Eq.~(\ref{eq:Mamp}) by the replacements
\begin{subequations}
\begin{eqnarray}
{\cal M}_\gamma
&\to& {\cal M}_\gamma + {\cal M}_{\gamma\gamma}\ ,	\\
{\cal M}_Z
&\to& {\cal M}_Z + {\cal M}_{\gamma Z} + {\cal M}_{Z \gamma}\ ,
\end{eqnarray}
\end{subequations}
where the two-photon and $\gamma Z$ exchange amplitudes
${\cal M}_{\gamma\gamma}$, ${\cal M}_{Z \gamma}$ and
${\cal M}_{\gamma Z}$ are given explicitly below.
The relative corrections from the $Z(\gamma\gamma)$, $\gamma(\gamma Z)$,
and $\gamma(\gamma\gamma)$ interference terms can be identified as
\begin{subequations}
\begin{eqnarray}
\delta_{Z (\gamma\gamma)}
&=& { 2\ \Re \left( {\cal M}_Z^* {\cal M}_{\gamma\gamma} \right)
    \over  2\ \Re \left( {\cal M}_Z^* {\cal M}_\gamma \right)}\ ,	\\
\delta_{\gamma (\gamma Z)}
&=& { 2\ \Re \left( {\cal M}_\gamma^* {\cal M}_{\gamma Z}
		  + {\cal M}_\gamma^* {\cal M}_{Z\gamma} \right)
    \over  2\ \Re \left( {\cal M}_\gamma^* {\cal M}_Z \right)}\ ,	\\
\delta_{\gamma (\gamma\gamma)}
&=& { 2\ \Re \left( {\cal M}_\gamma^* {\cal M}_{\gamma\gamma} \right)
    \over  | {\cal M}_\gamma |^2}\ .
\end{eqnarray}
\end{subequations}
The correction to the Born level PV asymmetry $A_{\rm PV}^0$ can then
be represented as
\be
A_{\rm PV}\ =\ (1+\delta) A_{\rm PV}^0\
\equiv\ \left({1 + \delta_{Z (\gamma\gamma)} + \delta_{\gamma(Z\gamma)}
	      \over 1 + \delta_{\gamma (\gamma\gamma)} }
	\right) A_{\rm PV}^0\ ,
\label{eq:delta_APV}
\ee
where $A_{\rm PV}$ is the full asymmetry, including TBE corrections,
and $A_{\rm PV}^0$ is given in Eq.~(\ref{eq:APV0}).
Since the electromagnetic TPE correction $\delta_{\gamma(\gamma\gamma)}$
is typically only a few percent \cite{BMT05,BMT03,KondD}, the full
correction $\delta$ can be written approximately as
\be
\delta \approx
\delta_{Z(\gamma\gamma)}
+ \delta_{\gamma(Z\gamma)}
- \delta_{\gamma(\gamma\gamma)}\ .
\label{eq:delta_approx}
\ee
In the model discussed here, the amplitudes ${\cal M}_{\gamma\gamma}$,
${\cal M}_{\gamma Z}$ and ${\cal M}_{Z \gamma}$ contain contributions
from both nucleon elastic and $\Delta(1232)$ isobar intermediate states,
which we discuss next.

\subsection{Nucleon Intermediate States}

For completeness, here we review the basic elements of the TBE
exchange calculation with nucleon intermediate states.
A more complete account can be found in Refs.~\cite{TM,BMT05,BMT03}.
For electromagnetic scattering, the total 2$\gamma$ exchange amplitude
for the box and crossed-box diagrams with a nucleon intermediate state
has the form \cite{BMT05}
\begin{eqnarray}
{\cal M}_{\gamma N \gamma}
&=& e^4 \int {d^4 k\over (2\pi)^4}
\ubar_e(p_3) \Big[ \gamma_\mu S_F(p_1-k,m_e) \gamma_\nu
                 + \gamma_\nu S_F(p_3+k,m_e) \gamma_\mu
             \Big] u_e(p_1)\              \nonumber \\
&\times&
\ubar_N(p_4)\ \Gamma_\gamma^\mu(q-k)\ S_F(p_2+k,M)\
          \Gamma_\gamma^\nu(k)\ u_N(p_2)\
\Delta_F(k,\lambda)\ \Delta_F(k-q,\lambda)\ ,
\label{eq:MgNg}
\end{eqnarray}
where $m_e$ is the electron mass, and the fermion (electron) and gauge
boson (photon) propagators are given by
\begin{eqnarray}
i S_F(k,m)
&=& { i\ (\fslash{k} + m) \over k^2 - m^2 + i\epsilon }\ ,  \\
i \Delta_F(k,\lambda)
&=& { -i \over k^2 - \lambda^2 + i\epsilon }\ ,
\end{eqnarray}
respectively, with $\lambda$ introduced as an infinitesimal photon mass
to regulate the infra-red divergences.

The calculation of the $\gamma$--$Z$ interference amplitude proceeds
along similar lines to that of the $2\gamma$ amplitudes above, with
the appropriate replacements of the photon propagator by the $Z$ boson
propagator, and the $\gamma N N$ vertex function by $\Gamma_Z^\mu$ in
Eq.~(\ref{eq:JZ}),
\begin{eqnarray}
{\cal M}_{\gamma N Z}
&=& {e^2 g^2 \over (4\, \cos\theta_W)^2}
\int {d^4 k\over (2\pi)^4}              \nonumber\\
& & \hspace*{-1cm} \times\
\ubar_e(p_3)
\Big[ (g_V^e \gamma_\mu + g_A^e \gamma_\mu \gamma_5)
       S_F(p_1-k,m) \gamma_\nu
    + \gamma_\nu S_F(p_3+k,m)
      (g_V^e \gamma_\mu + g_A^e \gamma_\mu \gamma_5)
\Big] u_e(p_1)\				\nonumber\\
& & \hspace*{-1cm} \times\
\ubar_N(p_4)\ \Gamma_Z^\mu(q-k)\ S_F(p_2+k,M)\
              \Gamma_\gamma^\nu(k)\ u_N(p_2)\
\Delta_F(k,\lambda)\ \Delta_F(k-q,M_Z)\ .
\label{eq:MgNZ}
\end{eqnarray}
A similar expression holds for the conjugate amplitude
${\cal M}_{Z N \gamma}$.

For the electromagnetic nucleon form factors we use the global fit to
the proton electric and magnetic form factors from Arrington {\em et al.}
\cite{AMT}, and for the neutron form factors from Bosted \cite{Bosted}.
For technical reasons, we parameterize the form factors by a sum
of three monopoles.
To examine the model dependence of the calculation, we also consider
a dipole shape for the proton form factors, with a dipole mass of
$\Lambda_{N (V)} = 0.84$~GeV \cite{BMT05,BMT03}.

The weak $ZNN$ form factors are less well determined.
Using the conservation of the vector current (CVC), the weak vector
form factors can be directly related to the $\gamma NN$ form factors.
For the axial-vector form factor, on the other hand, we use an
empirical dipole fit, $G_A(Q^2) = G_A(0)/(1+Q^2/\Lambda_{N (A)}^2)^2$,
where $G_A(0)=1.267$ is the axial vector charge, with the mass
parameter $\Lambda_{N (A)} = 1$~GeV.
Varying $\Lambda_{N (A)}$ by 20\% does not affect the results
significantly.
Since the main purpose of the PV experiments is to extract strange
quark contributions to form factors by comparing the measured asymmetry
with the predicted zero-strangeness asymmetry, in all our numerical
simulations we set the strange form factors to zero,
$F_{1,2}^s = 0 = G_A^s$.

\begin{figure}
\includegraphics[height=5.5cm]{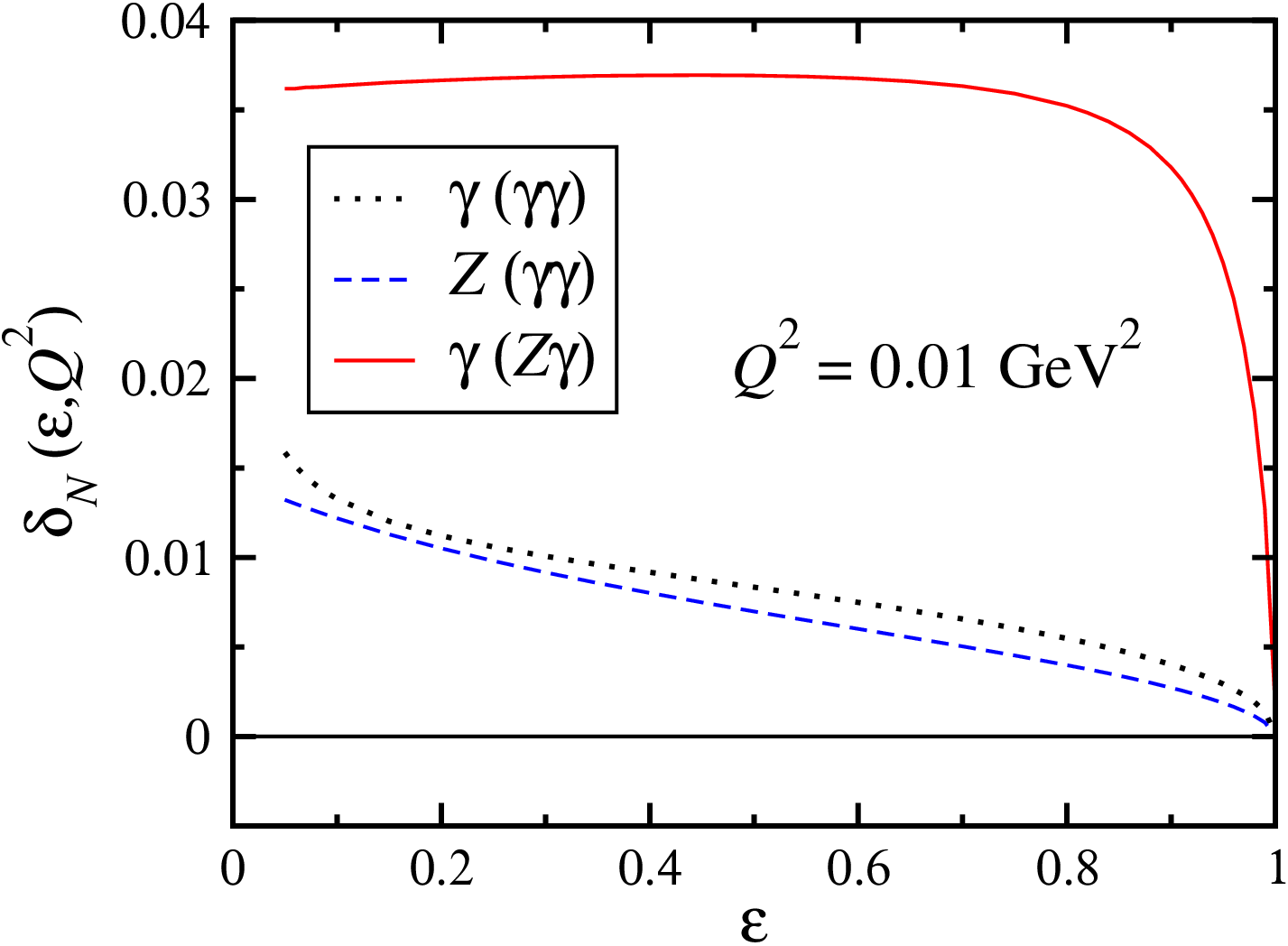}\hspace*{0.5cm}
\includegraphics[height=5.5cm]{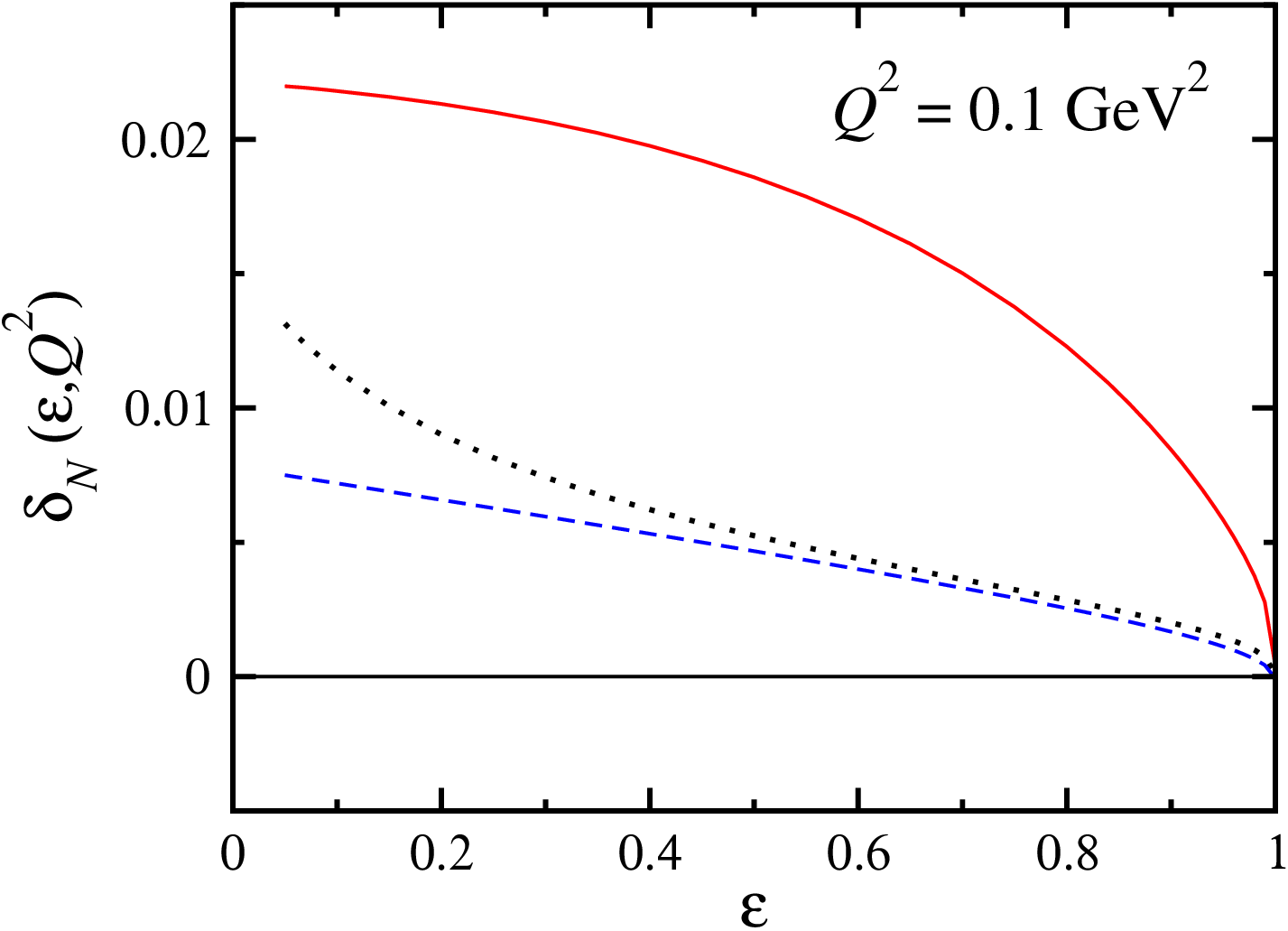}\vspace*{1cm}
\includegraphics[height=5.5cm]{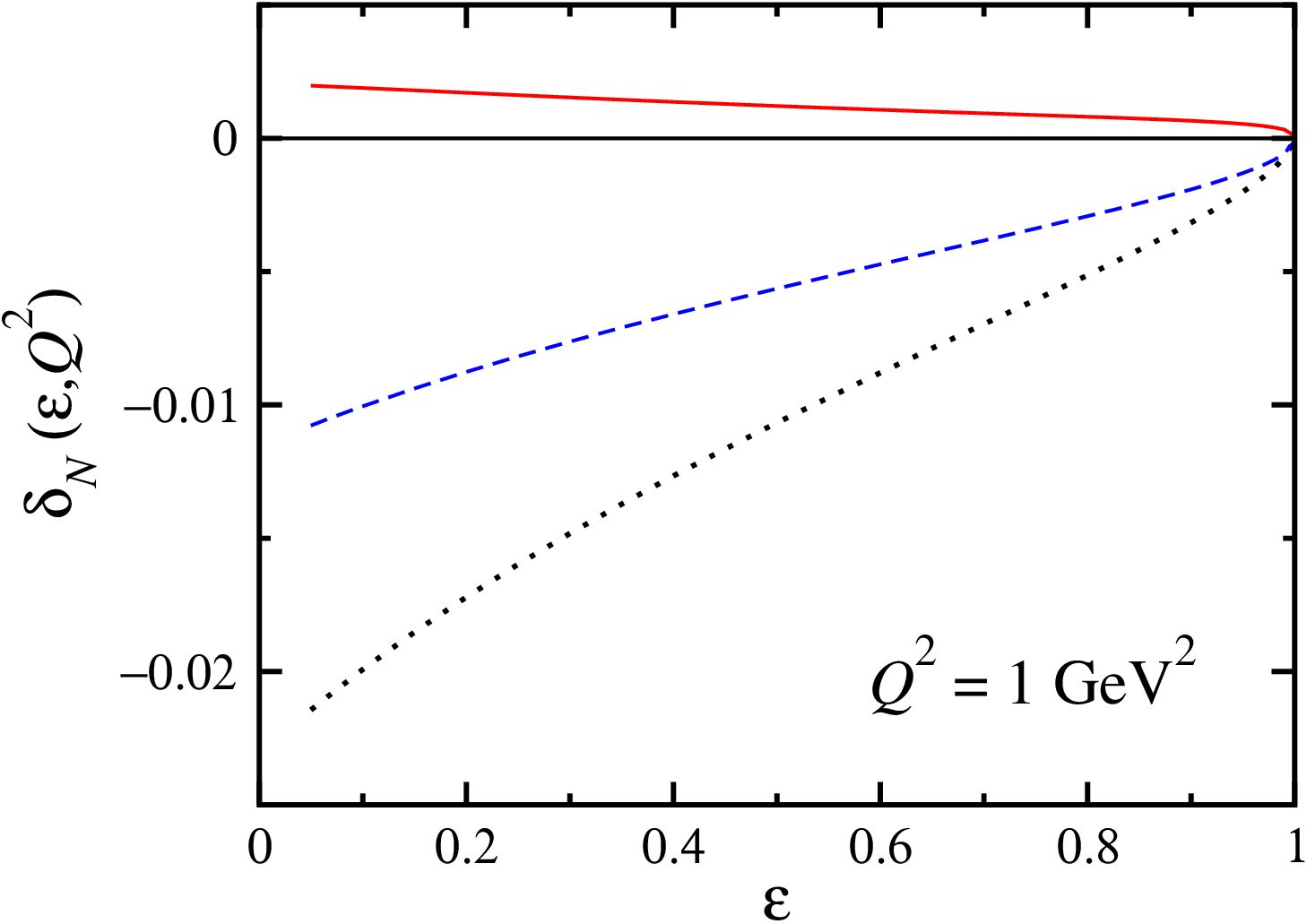}\hspace*{0.5cm}
\includegraphics[height=5.5cm]{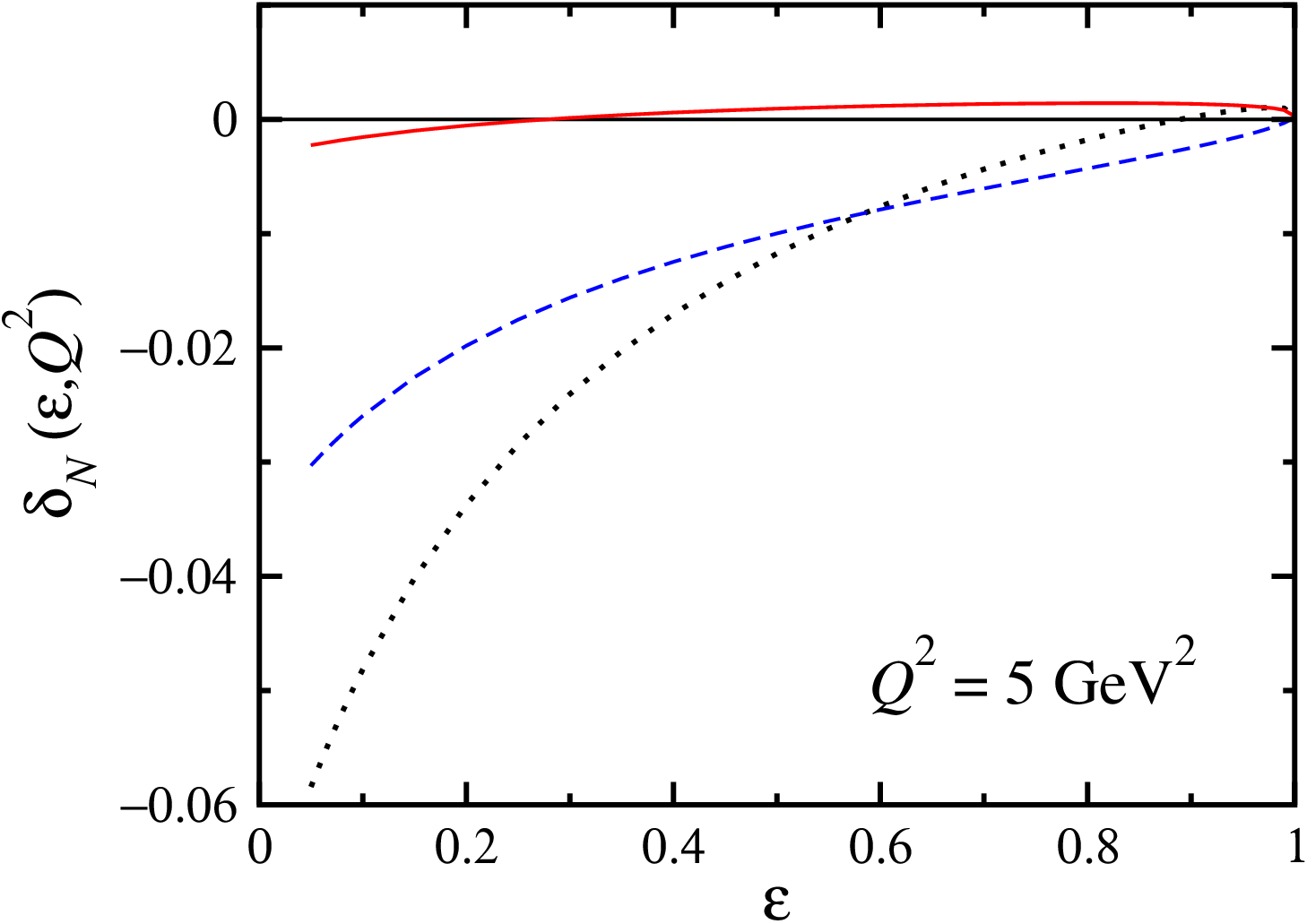}
\caption{TBE corrections $\delta_N(\eps,Q^2)$ with nucleon
    intermediate states, for the $\gamma(\gamma\gamma)$ (dotted),
    $Z(\gamma\gamma)$ (dashed) and $\gamma(Z\gamma)$ (solid)
    contributions at $Q^2=0.01$, 0.1, 1 and 5~GeV$^2$.
    The correction is defined relative to that of Mo \& Tsai \cite{MT}.
    Note that the $\gamma(\gamma\gamma)$ correction enters with
    the opposite sign in the asymmetry, Eq.~(\ref{eq:delta_approx}).
}
\label{fig:delN}
\end{figure}

In Fig.~\ref{fig:delN} we show the various contributions to the
two-boson exchange correction $\delta_N$ as a function of $\eps$
for several values of $Q^2$ ($Q^2 = 0.01$, 0.1, 1 and 5~GeV$^2$).
The infrared divergences \cite{MT,MTj} in the boxes have been
removed following the standard treatment of Mo \& Tsai \cite{MT}.
It should be noted, however, that, in contrast to the ${2\gamma}$ box
diagrams, the infrared contributions for the $\gamma Z$ box diagrams
are significantly different using the procedure of Ref.~\cite{MTj}.
At small $Q^2$ values ($Q^2 \lesssim 0.1$~GeV$^2$) the
$\gamma (\gamma \gamma)$ and $Z (\gamma \gamma)$ contributions
are very similar, and considerably smaller in magnitude than the
$\gamma (Z \gamma)$ component.
Since the $\gamma$--$Z$ interference and the purely electromagnetic
contributions enter in the numerator and denominator of the PV
asymmetry, respectively, the $\gamma (\gamma \gamma)$ and
$Z (\gamma \gamma)$ will partially cancel in their effect on
$A_{\rm PV}$, which will be determined mostly by the
$\gamma (Z \gamma)$ component.
At larger $Q^2$ ($\gtsim 1$~GeV$^2$) the $\gamma (Z \gamma)$
component decreases in magnitude, while the $\gamma (\gamma \gamma)$
$Z (\gamma \gamma)$ pieces become large and more negative
\cite{TM,BMT05,BMT03}.

\begin{figure}
\includegraphics[height=7cm]{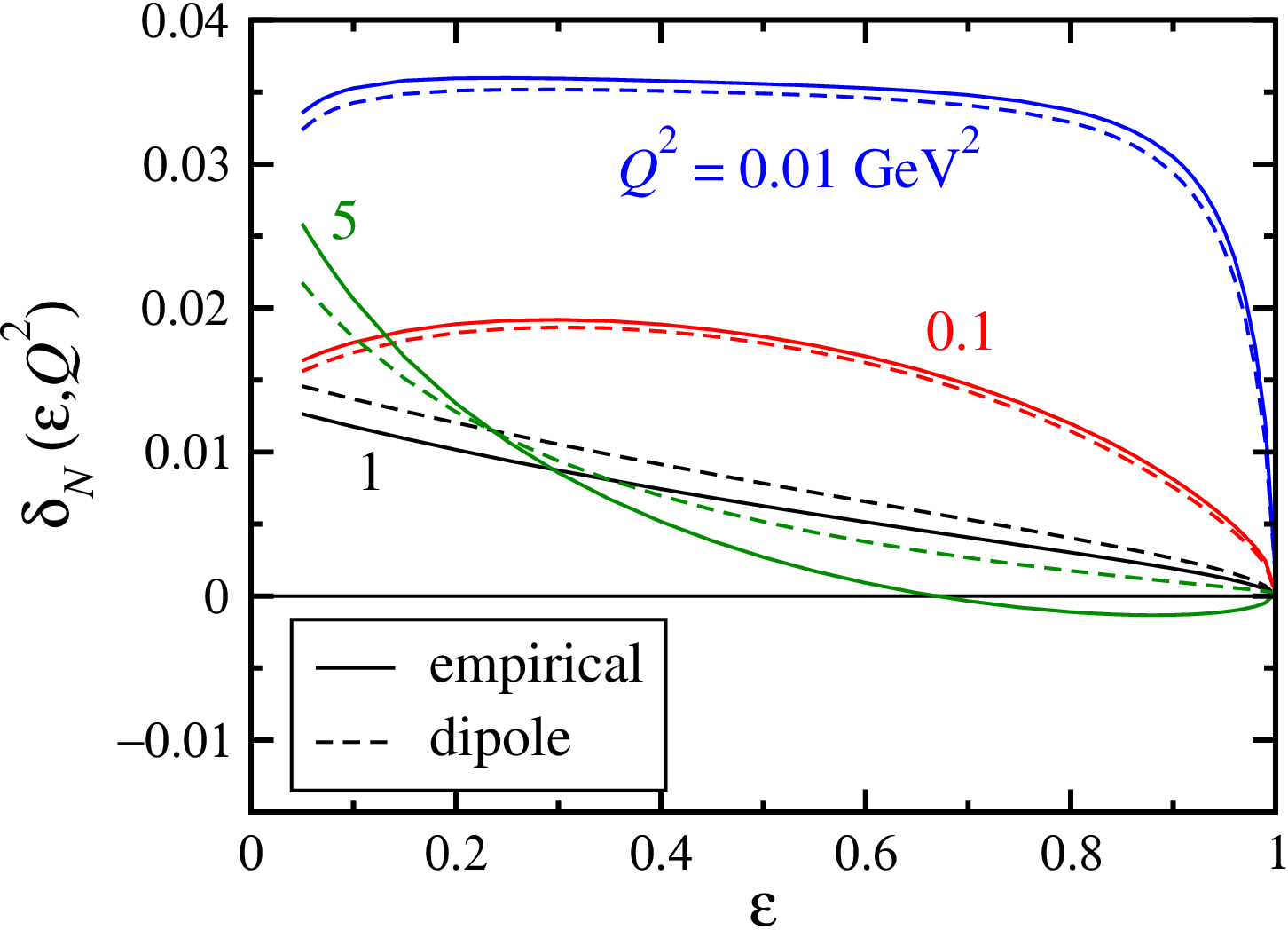}
\caption{Model dependence of the total TBE corrections
	$\delta_N(\eps,Q^2)$ with nucleon intermediate
	states for $Q^2=0.01$, 0.1, 1 and 5~GeV$^2$, using the
	empirical form factors as described in the text (solid),
	and using a dipole approximation (dashed).}
\label{fig:delN_model}
\end{figure}

The dependence of the total correction $\delta_N$ on the input form
factors is illustrated in Fig.~\ref{fig:delN_model}.
The difference between the results using the empirical form factors
and the dipole approximation is very small for all values of $\eps$,
and only becomes appreciable at large $Q^2$ ($Q^2 \gtsim 1$~GeV$^2$),
consistent with the findings of our earlier analysis \cite{BMT05}.
Interestingly, the correction at $Q^2 = 0.01$~GeV$^2$ is relatively
flat over the range $0.1 \lesssim \eps \lesssim 0.8$, before dropping
rapidly as $\eps \to 1$.
At large $Q^2$ the total TBE correction becomes more strongly $\eps$
dependent, decreasing in magnitude at forward scattering angles but
increasing at backward angles ($\eps \to 0$).

\subsection{$\Delta$ Intermediate States}

In evaluating the contribution to the TBE amplitude from the
excitation of the $\Delta(1232)$-isobar, we use the formalism
outlined in Ref.~\cite{KondD} for the $\gamma N\Delta$ interaction,
and extend this to the weak sector with the introduction of axial
$ZN\Delta$ couplings.
The $\gamma N\Delta$ vertex is given by \cite{KondD,KS}
\begin{eqnarray}
\Gamma_{\gamma\Delta \to N}^{\mu\alpha}(p,q)
&=& {i \over 2 M_\Delta^2}\sqrt{2\over 3}
\Big\{
  g_1(Q^2)
  \left[ g^{\mu\alpha} \fslash{p}\ \fslash{q}
    - p^\mu \gamma^\alpha \fslash{q}
    - \gamma^\mu \gamma^\alpha p\cdot q
    + \gamma^\mu q^\alpha \fslash{p}
  \right]                   \nonumber\\
& & \hspace*{-2cm}
+\ g_2(Q^2)
  \left[ p^\mu q^\alpha - g^{\mu\alpha} p\cdot q
  \right]\
+\ {g_3(Q^2) \over M_\Delta}
  \left[ q^2 \left( p^\mu \gamma^\alpha - g^{\mu\alpha} \fslash{p} \right)
       + q^\mu \left( q^\alpha \fslash{p} - \gamma^\alpha p\cdot q \right)
  \right]
\Big\}\ \gamma_5\ ,             \nonumber\\
& &
\label{eq:gDN}
\end{eqnarray}
where $p$ and $q$ are the incoming $\Delta$ and photon momenta,
with corresponding Lorentz indices $\alpha$ and $\mu$, respectively.
The overall factor $\sqrt{2/3}$ arises from the $N \to \Delta$
isospin transition operator.
%
%
Electromagnetic gauge invariance implies that
$q_\mu \Gamma_{\gamma\Delta \to N}^{\mu\alpha}(p,q) = 0$.
The coupling constants s $g_i \equiv g_i(Q^2=0)$ for $i=1,2,3$ can
be related to the magnetic, electric and Coulomb components of the
$\gamma N\Delta$ vertex by $g_1 = g_M$, $g_E = g_2-g_1$, $g_C = g_3$.
The vertex with an outgoing $\Delta$ can be obtained from the relation
\be
\Gamma_{\gamma N \to \Delta}^{\alpha\mu}(p,q)
= \gamma_0 \left[ \Gamma_{\gamma\Delta \to N}^{\mu\alpha}(p,q)
       \right]^\dagger \gamma_0\ ,
\ee
where $p$ is the outgoing $\Delta$ momentum and $q$ the incoming
photon momentum.

The amplitude for the box and crossed-box diagrams with a $\Delta$
intermediate state can then be written as
\begin{eqnarray}
{\cal M}_{\gamma\Delta\gamma}
&=& e^4 \int {d^4 k\over (2\pi)^4}
\ubar_e(p_3)
  \Big[ \gamma_\mu S_F(p_1-k,m_e) \gamma_\nu
      + \gamma_\nu S_F(p_3+k,m_e) \gamma_\mu
  \Big] u_e(p_1)\      				\nonumber\\
& & \hspace*{-1cm} \times\
\ubar_N(p_4)\
  \Gamma_{\gamma\Delta\to N}^{\mu\alpha}(p_2+k,q-k)\
  S_F(p_2+k,M)\ {\cal P}^{3/2}_{\alpha\beta}(p_2+k)\
  \Gamma_{\gamma N\to\Delta}^{\beta\nu}(p_2+k,k)\
u_N(p_2)          \nonumber\\
& & \hspace*{-1cm} \times\
\Delta_F(k,0)\ \Delta_F(k-q,0)\ ,
\label{eq:MgDg}
\end{eqnarray}
where the projection operator
\be
{\cal P}^{3/2}_{\alpha\beta}(p)
= g_{\alpha\beta}
- {1 \over 3} \gamma_\alpha \gamma_\beta
- {1 \over 3 p^2} (\fslash{p} \gamma_\alpha p_\beta
          + p_\alpha \gamma_\beta \fslash{p})
\ee
ensures that only spin-3/2 components are present.
Suppression of the unphysical spin-1/2 contributions also leads to
the condition on the vertex
$p_\alpha \Gamma_{\gamma\Delta \to N}^{\mu\alpha}(p,q) = 0$.
Note that in Eq.~(\ref{eq:MgDg}) a finite photon mass is not needed
in the photon propagators, since, in contrast to Eq.~(\ref{eq:MgNg}),
the result here is infra-red finite.

For simplicity, we assume a dipole shape for the three
$\gamma N\Delta$ transition form factors,
$g_i(Q^2) \equiv g_i\ F_V^\Delta(Q^2)$ for $i=1,2,3$, 
where $F_V^\Delta(Q^2) = (1+Q^2/\Lambda_{\Delta (V)}^2)^{-2}$,
with a dipole mass $\Lambda_{\Delta (V)} = 0.84$~GeV for each.
For the electric and magnetic couplings we use the values $g_1=7$
and $g_2=9$ \cite{KondD}, obtained from a K-matrix analysis of pion
photoproduction data \cite{KS}.
A more realistic $\pi N$ coupled channel quasi-potential study
\cite{PT04} gives similar values, $g_1=6.3$ and $g_2=9.7$.
For the $g_3$ coupling, an estimate from the $\gamma N \to \Delta$
E2/M1 transition strength yields $g_3=5.8$.
To test the sensitivity of the TBE corrections to the value of $g_3$,
we consider a range of couplings, as discussed below.
Note that the interference contributions between the $g_1$, $g_2$
and $g_3$ terms cancel in the TBE amplitude because of the odd and
even character of these vertices in the loop variable $k$.

For the $ZN\Delta$ vertex both vector and axial-vector contributions
enter.
For the vector transitions, CVC requires the same form for the
$ZN\Delta$ vertex as for the $\gamma N\Delta$,
\begin{eqnarray}
\Gamma_{Z\Delta \to N}^{\mu\alpha (V)}(p,q)
&=& {i \over 2 M_\Delta^2}\sqrt{2\over 3}
\Big\{
  g_1^V(Q^2)
  \left[ g^{\mu\alpha} \fslash{p}\ \fslash{q}
    - p^\mu \gamma^\alpha \fslash{q}
    - \gamma^\mu \gamma^\alpha p\cdot q
    + \gamma^\mu q^\alpha \fslash{p}
  \right]                   \nonumber\\
& & \hspace*{-2cm}
+\ g_2^V(Q^2)
  \left[ p^\mu q^\alpha - g^{\mu\alpha} p\cdot q
  \right]
+\ {g_3^V(Q^2) \over M_\Delta}
  \left[ q^2 \left( p^\mu \gamma^\alpha - g^{\mu\alpha} \fslash{p} \right)
       - q^\mu \left( q^\alpha \fslash{p} - \gamma^\alpha p\cdot q \right)
  \right]
\Big\}\ \gamma_5\ ,          \nonumber\\
& &
\label{eq:ZDN_V}
\end{eqnarray}
where again the factor $\sqrt{2/3}$ is associated with the $N \to \Delta$
weak isospin transition.
Using CVC and isospin symmetry, the vector $ZN\Delta$ form factors
can be related to the $\gamma N \Delta$ form factors by
\be
g_i^V(Q^2) = 2 (1 - 2\sw)\ g_i(Q^2)\ ,
\ee
where the $Q^2$ dependence of the electromagnetic $\gamma N \Delta$
form factor is parameterized as above.

For the axial-vector vertex, nonconservation of the axial current
implies the existence of an addition form factor.
However, one can use the partially conserved axial current (PCAC)
hypothesis to relate two of the form factors, leaving a similar
expression to that in Eq.~(\ref{eq:ZDN_V}),
\begin{eqnarray}
\Gamma_{Z\Delta \to N}^{\mu\alpha (A)}(p,q)
&=& {i \over 2 M_\Delta^2}
\Big\{
  g_1^A(Q^2)
  \left[ g^{\mu\alpha} \fslash{p}\ \fslash{q}
        - p^\mu \gamma^\alpha \fslash{q}
        - \gamma^\mu \gamma^\alpha p\cdot q
        + \gamma^\mu q^\alpha \fslash{p}
  \right]                   \nonumber\\
& & \hspace*{-2cm}
+\ g_2^A(Q^2)
  \left[ p^\mu q^\alpha - g^{\mu\alpha} p\cdot q
  \right]\
+\ {g_3^A(Q^2) \over M_\Delta}
  \left[ q^2 \left( p^\mu \gamma^\alpha - g^{\mu\alpha} \fslash{p} \right)
       - q^\mu \left( q^\alpha \fslash{p} - \gamma^\alpha p\cdot q \right)
  \right]
\Big\} .                  \nonumber\\
& &
\label{eq:ZDN_A}
\end{eqnarray}
Note that here the weak isospin transition factor has been absorbed
into the definition of the couplings \cite{Paschos}.
The axial form factors are less well determined, but some constraints
have been extracted from analysis of $\nu$ scattering data.
In a recent analysis, Lalakulich \& Paschos \cite{Paschos}
parameterized the $\nu N \to \mu \Delta$ cross sections from bubble
chamber experiments at low $Q^2$ in terms of phenomenological form
factors.
The available data can be described by the form factors
$g_1^A(Q^2) = 0$,
$g_2^A(Q^2) = (M_\Delta^2/2 M^2)\, C_5^A(Q^2)
	    = (Q^2/4 M^2)\, g_3^A(Q^2)$,
where $C_5^A$ is given in Appendix~A, with $C_5^A(Q^2=0) = 1.2$
\cite{Paschos}.
For the $Q^2$ dependence we again take a dipole form, with a cut-off
mass of $\Lambda_{\Delta (A)} = 1.0$~GeV.

As for the electromagnetic case, the vertex with an outgoing $\Delta$
can be obtained from the relation
\be
\Gamma_{Z N \to \Delta}^{\alpha\mu (V,A)}(p,q)
= \gamma_0 \left[ \Gamma_{Z \Delta \to N}^{\mu\alpha (V,A)}(p,q)
       \right]^\dagger \gamma_0\ ,
\ee
where $p$ is the outgoing $\Delta$ momentum and $q$ the incoming
$Z$-boson momentum.
The $ZN\Delta$ amplitude for the box and crossed-box diagrams with
a $\Delta$ intermediate state can then be written
\begin{eqnarray}
{\cal M}_{\gamma\Delta Z}
&=& {e^2 g^2 \over (4 \cos\theta_W)^2}
\int {d^4 k\over (2\pi)^4}              \nonumber\\
& & \hspace*{-1cm} \times\
\ubar_e(p_3)
\Big[
  (g_V^e \gamma_\mu + g_A^e \gamma_\mu \gamma_5)
  S_F(p_1-k,m) \gamma_\nu
+ \gamma_\nu S_F(p_3+k,m)
  (g_V^e \gamma_\mu + g_A^e \gamma_\mu \gamma_5)
\Big] u_e(p_1)\                          \nonumber\\
& & \hspace*{-1cm} \times\
\ubar_N(p_4) \Gamma_{Z\Delta\to N}^{\mu\alpha}(p_2+k,q-k)
         S_F(p_2+k,M)\ {\cal P}^{3/2}_{\alpha\beta}(p_2+k)
                            \nonumber\\
& & \hspace*{-1cm} \times\
         \Gamma_{\gamma N\to\Delta}^{\beta\nu}(p_2+k,k)
         u_N(p_2)\
\Delta_F(k,0)\ \Delta_F(k-q,M_Z)\ ,
\label{eq:MgDZ}
\end{eqnarray}
where $\Gamma_{Z\Delta\to N}^{\mu\alpha}$ is the sum of the vector
(\ref{eq:ZDN_V}) and axial-vector (\ref{eq:ZDN_A}) vertices.
The corresponding amplitude ${\cal M}_{\gamma\Delta Z}$ can be
derived in a similar manner.

\begin{figure}
\includegraphics[height=5.5cm]{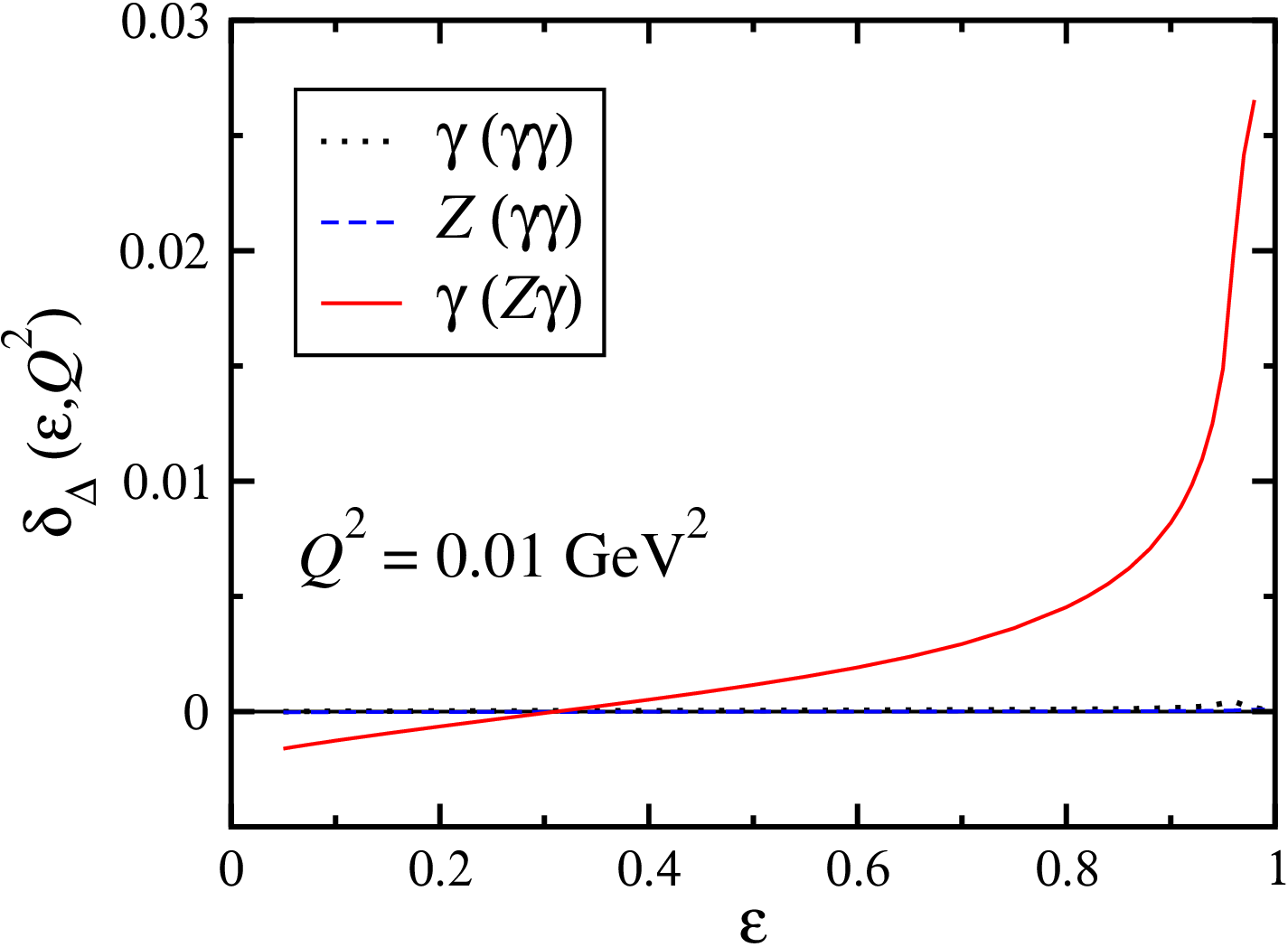}\hspace*{0.5cm}
\includegraphics[height=5.5cm]{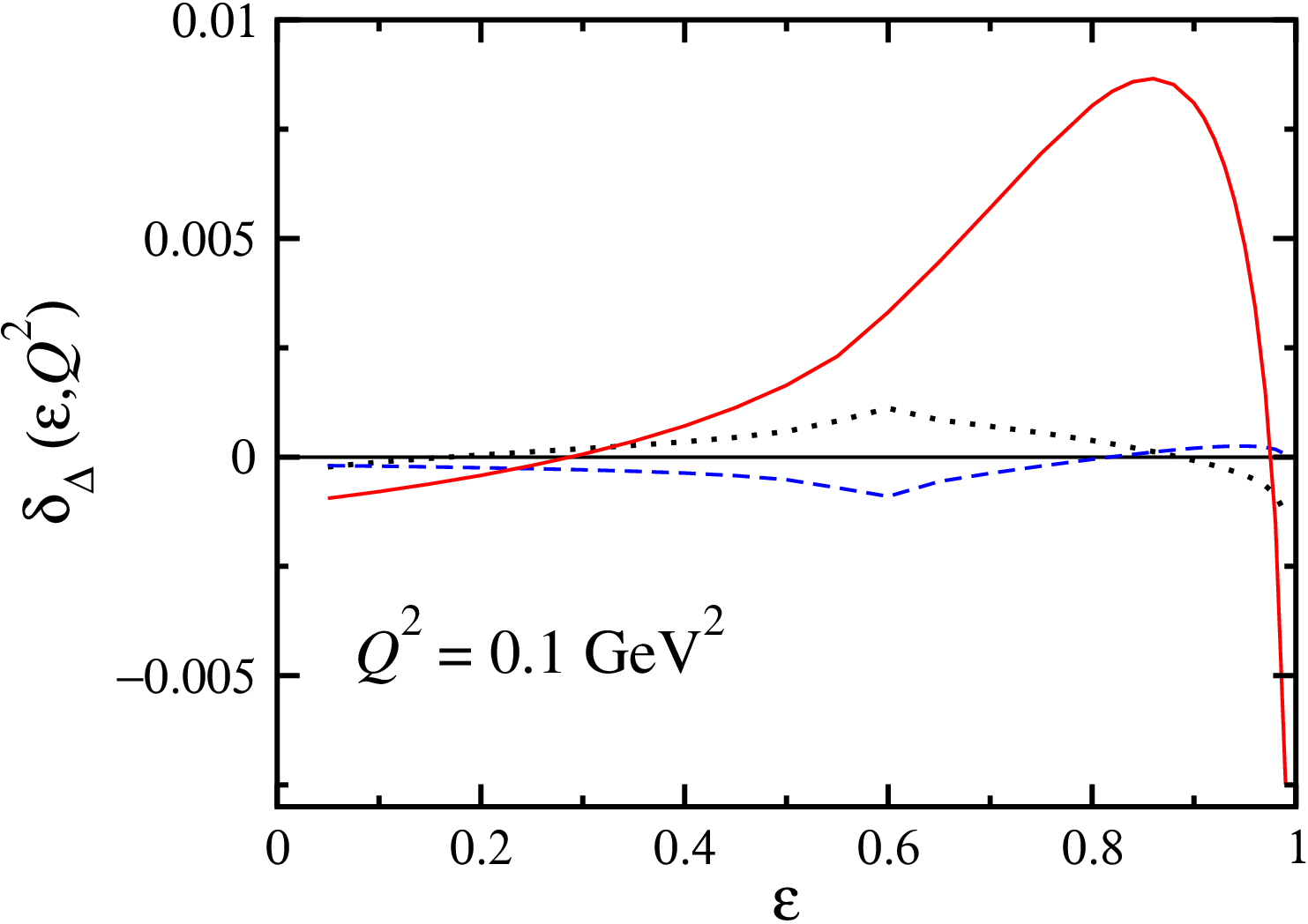}\vspace*{1cm}
\includegraphics[height=5.5cm]{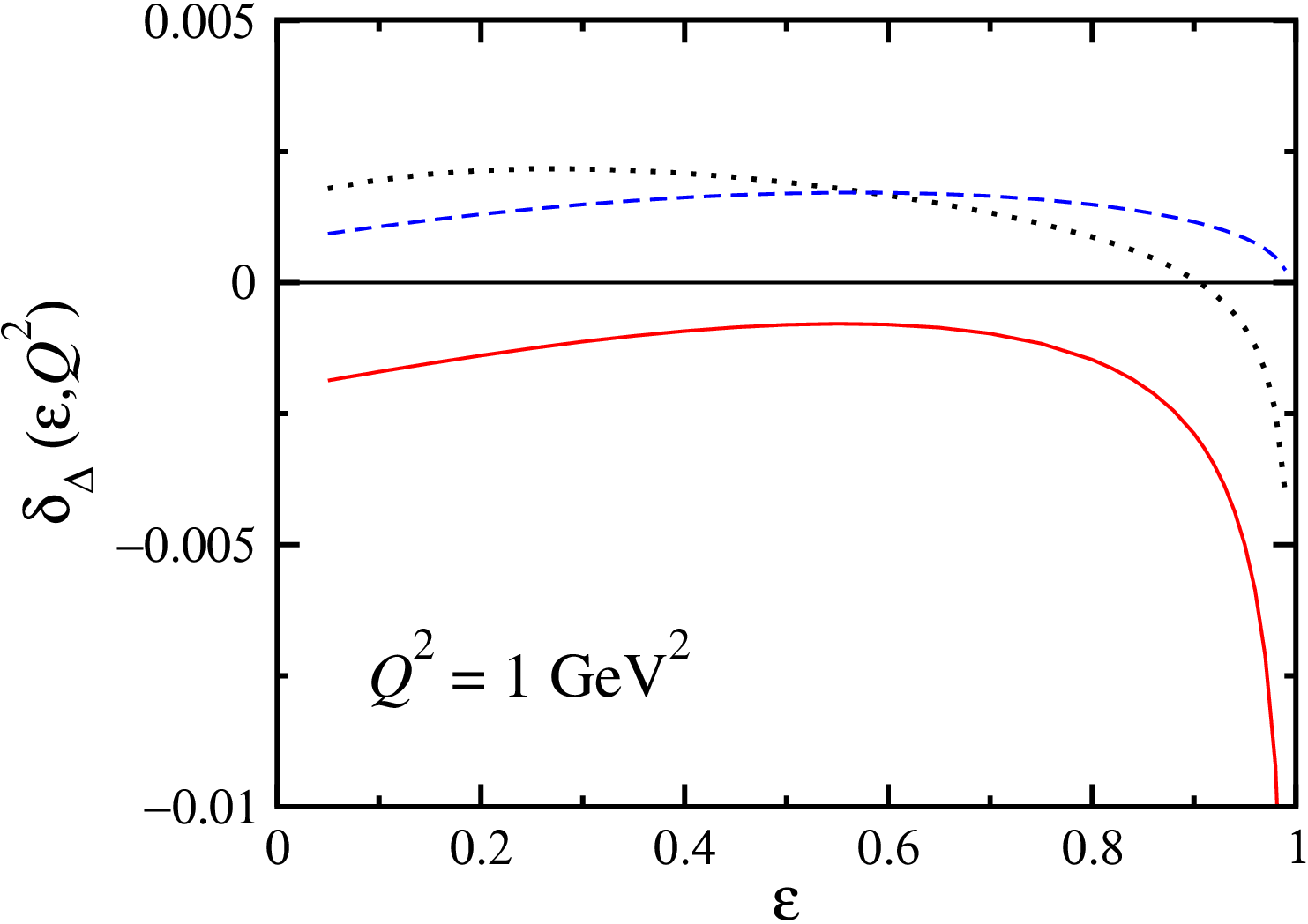}\hspace*{0.5cm}
\includegraphics[height=5.5cm]{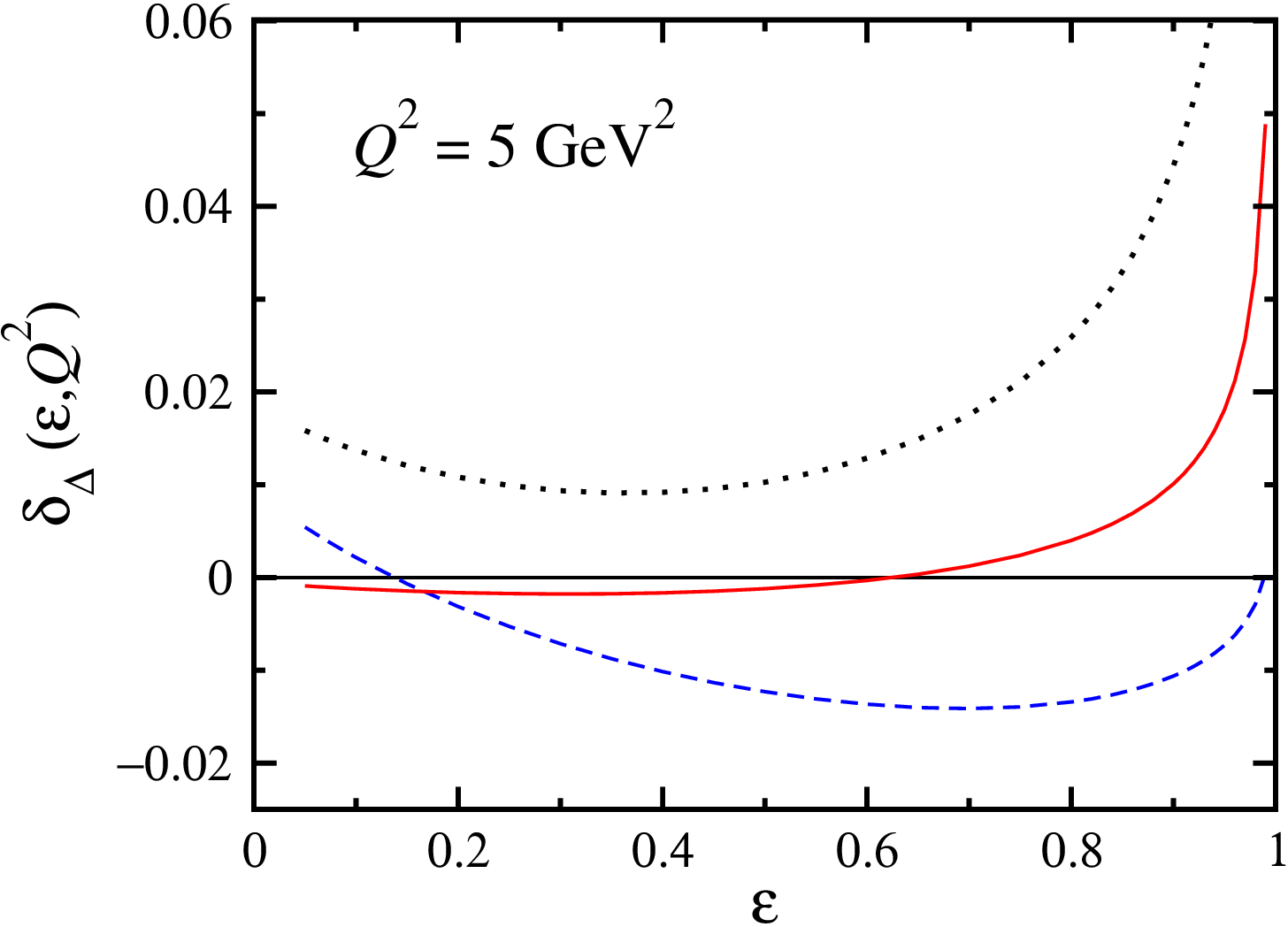}
\caption{TBE corrections $\delta_\Delta(\eps,Q^2)$
	with $\Delta(1232)$ intermediate states, for the
	$\gamma(\gamma\gamma)$ (dotted), $Z(\gamma\gamma)$
	(dashed) and $\gamma(Z\gamma)$ (solid)
	contributions at $Q^2=0.01$, 0.1, 1 and 5~GeV$^2$.}
\label{fig:delD}
\end{figure}

In Fig.~\ref{fig:delD} we plot the individual TBE contributions to
$\delta_\Delta$ from processes with intermediate $\Delta(1232)$
states as a function of $\eps$ for a range of $Q^2$ values between
0.01 and 5~GeV$^2$.
Several interesting features can be noted.
Firstly, the magnitude and shape of the $\Delta$ corrections are
very different to the nucleon corrections in Fig.~\ref{fig:delN}.
At low $Q^2$ ($\lesssim 0.1$~GeV$^2$) the two-photon interference
with either the Born $\gamma$ or $Z$ exchange is almost negligible,
increasing somewhat at larger $Q^2$.
The $\gamma (Z \gamma)$ contribution is also relatively small at
low $\eps$, and none of the corrections exceed $\sim 1\%$ in magnitude
for $\eps \lesssim 0.8$ and $Q^2 \lesssim 1$~GeV$^2$, and $\sim 2\%$
for $Q^2 \lesssim 5$~GeV$^2$.

At larger $\eps$, however, the $\gamma (Z \gamma)$ correction
increases rapidly, becoming even bigger than the nucleon correction,
and in fact appears to diverge as $\eps \to 1$.
The increase of the one-loop contributions to the asymmetries may be
related to the growth of the invariant center of mass energy for fixed
$Q^2$ as $\eps \to 1$.
Since the $\Delta$ intermediate state amplitudes
${\cal M}_{\gamma \Delta \gamma}$ and ${\cal M}_{\gamma \Delta Z}$
have numerators which have higher powers of loop momenta than the
corresponding nucleon amplitudes
${\cal M}_{\gamma N \gamma}$ and ${\cal M}_{\gamma N Z}$,
one expects that the $\Delta$ contributions should grow faster with
invariant energy than the nucleon.
It is also interesting to observe the cusp behavior of the
$\gamma (\gamma \gamma)$ and $Z (\gamma \gamma)$ corrections at
$Q^2=0.1$~GeV$^2$ around $\eps=0.6$, the kinematics of which
corresponds to the threshold point of the $e$--$\Delta$ channel.

\begin{figure}[h]
\vspace*{1cm}
\includegraphics[height=5.5cm]{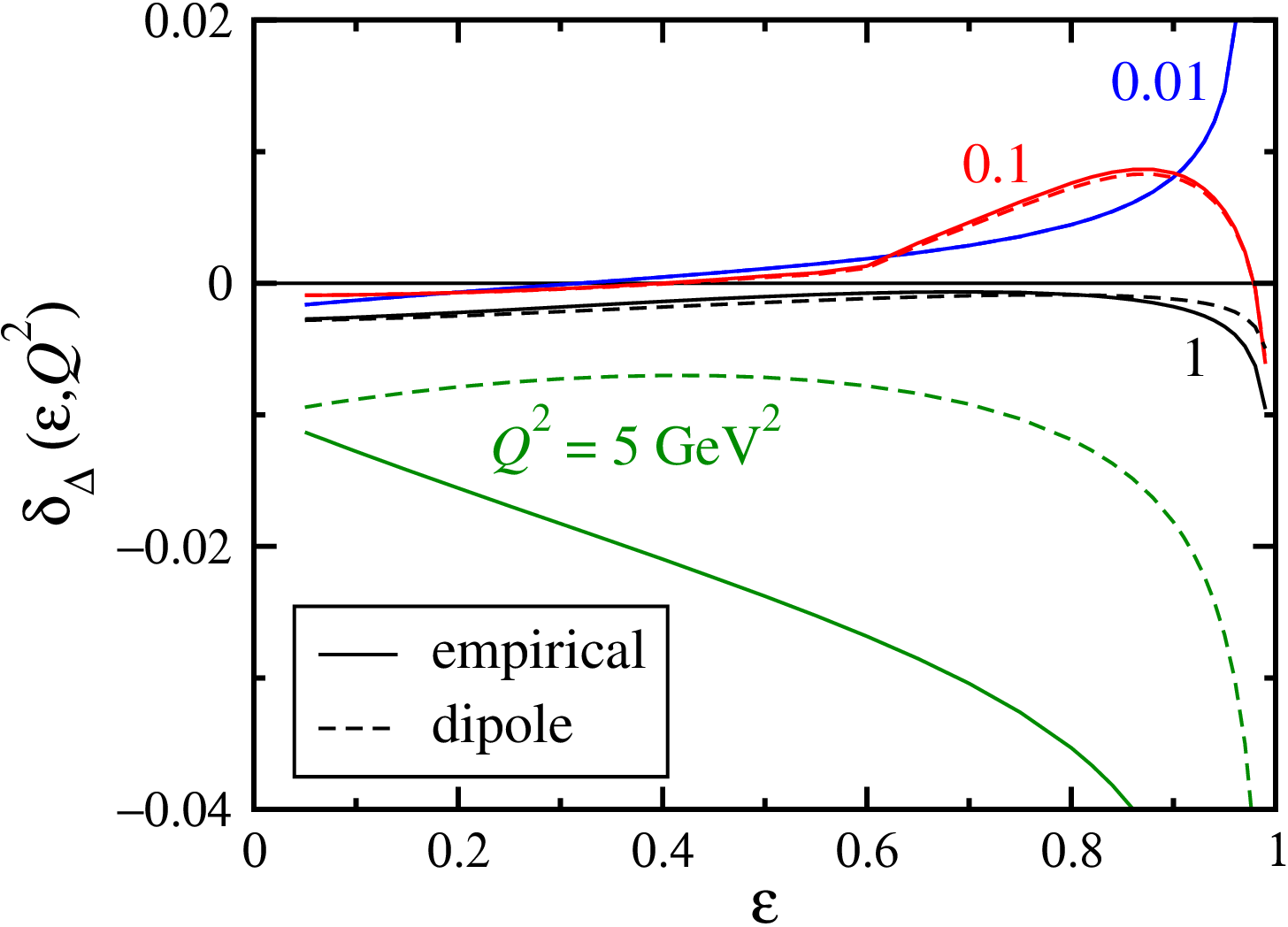}\hspace*{0.5cm}
\includegraphics[height=5.5cm]{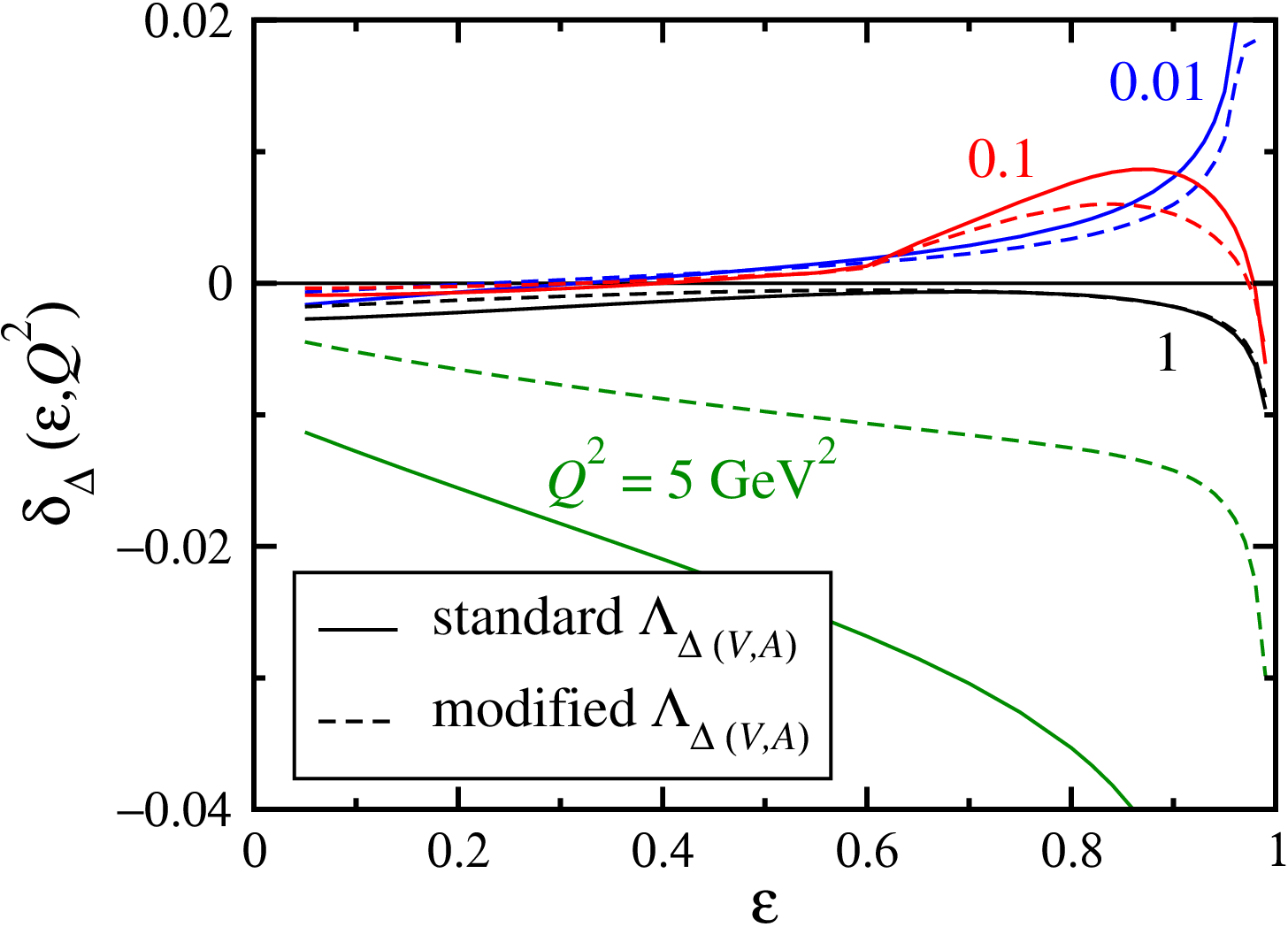}
\caption{Total TBE correction $\delta_\Delta(\eps,Q^2)$
	with $\Delta(1232)$ intermediate states for
	$Q^2=0.01$, 0.1, 1 and 5~GeV$^2$.
	(a) Comparison between using empirical nucleon form factors
	(solid) and a dipole approximation (dashed).
	(b) Dependence on the $N \to \Delta$ transition form factors,
	using the standard cut-offs
	$\Lambda_{\Delta (V)} = 0.84$~GeV,
	$\Lambda_{\Delta (A)} = 1.0$~GeV
	as described in the text (solid), and the modified cut-offs
	$\Lambda_{\Delta (V)} = 0.68$~GeV,
        $\Lambda_{\Delta (A)} = 0.8$~GeV (dashed).\\}
\label{fig:delD_model}
\end{figure}

The combined TBE correction from $\Delta$ intermediate states is
shown in Fig.~\ref{fig:delD_model}(a), for various input form factors.
In general the behavior of the total correction $\delta_\Delta$
is quite dramatic at high $\eps$, with the magnitude increasing
as $\eps \to 1$.
The total correction for $Q^2 \lesssim 0.1$~GeV$^2$ is positive
for most $\eps$ values, but changes sign to become negative at
larger $Q^2$.
As for the nucleon case, the dependence on the input form factors
is relatively weak for all $Q^2 \lesssim 1$~GeV$^2$, whether one
uses empirical form factors for the vector $\gamma NN$ or $ZNN$
vertices or a dipole approximation for all the form factors.
Similarly, the dependence on the dipole cut-off masses
$\Lambda_{\Delta (V,A)}$ for the $\gamma N\Delta$ and $ZN\Delta$
vertices is small for the same $Q^2$ range,
Fig.~\ref{fig:delD_model}(b). 
The sensitivity to the input form factors becomes more appreciable at
larger $Q^2$, however, as the $Q^2 = 5$~GeV$^2$ results demonstrate.
One should caution, though, that at momentum transfers of
$Q^2 \sim 5$~GeV$^2$ or higher the reliability of a purely hadronic
resonance description of the TBE process is more questionable.

\begin{figure}[h]
\includegraphics[height=6cm]{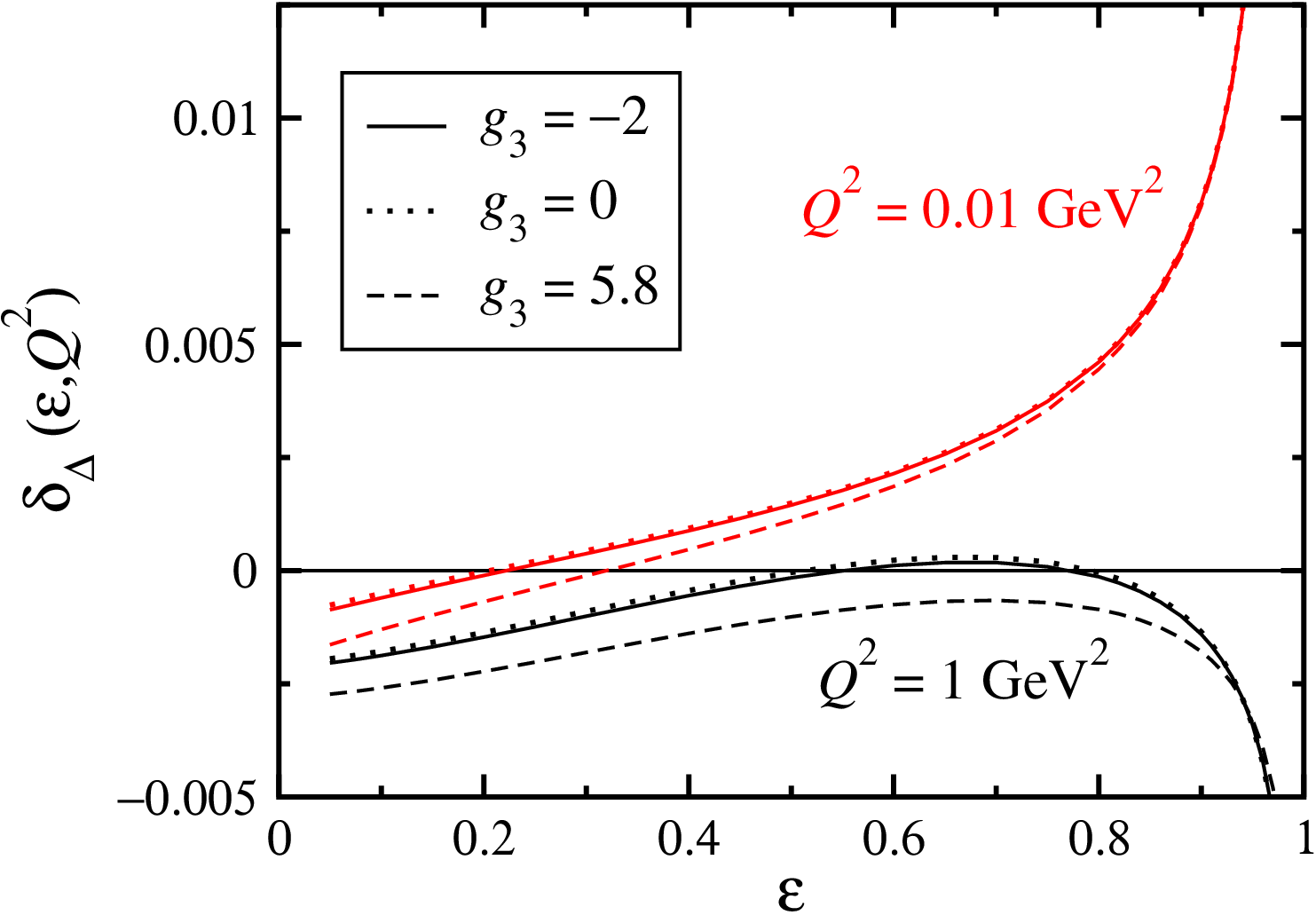}
\caption{Total TBE correction $\delta_\Delta(\eps,Q^2)$
	with $\Delta(1232)$ intermediate states for
	$Q^2=0.01$ and 1~GeV$^2$, with different Coulomb
	couplings $g_3 = -2$ (solid), 0 (dotted) and
	5.8 (dashed).}
\label{fig:delD_g3}
\end{figure}

Finally, the dependence of $\delta_\Delta$ on the Coulomb coupling
constant $g_3$ is illustrated in Fig.~\ref{fig:delD_g3}, where the
total correction at $Q^2=0.01$ and 1~GeV$^2$ is shown for $g_3 = -2$
\cite{KondD}, 0 and 5.8 \cite{KS}.
The results with $g_3 = -2$ and 0 are almost indistinguishable,
while using the preferred coupling $g_3 = 5.8$ gives slightly
smaller contributions for most $\eps$.
One can conclude, therefore, that the uncertainty in the Coulomb
coupling should not affect the overall results or conclusions.

\newpage

\newpage
\section{Effects on Observables}
\label{sec:res}

\begin{figure}
\includegraphics[height=5.5cm]{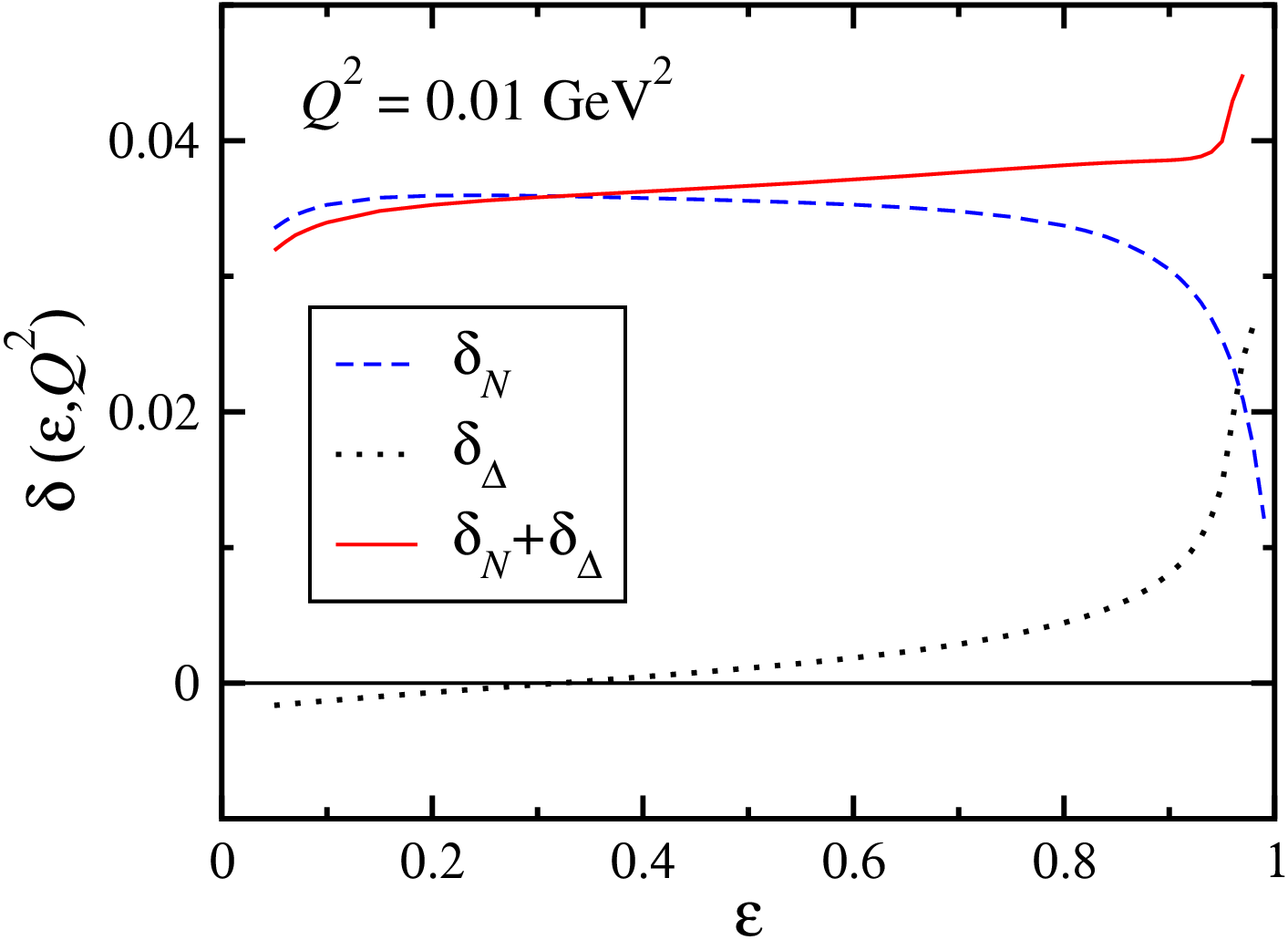}\hspace*{0.5cm}
\includegraphics[height=5.5cm]{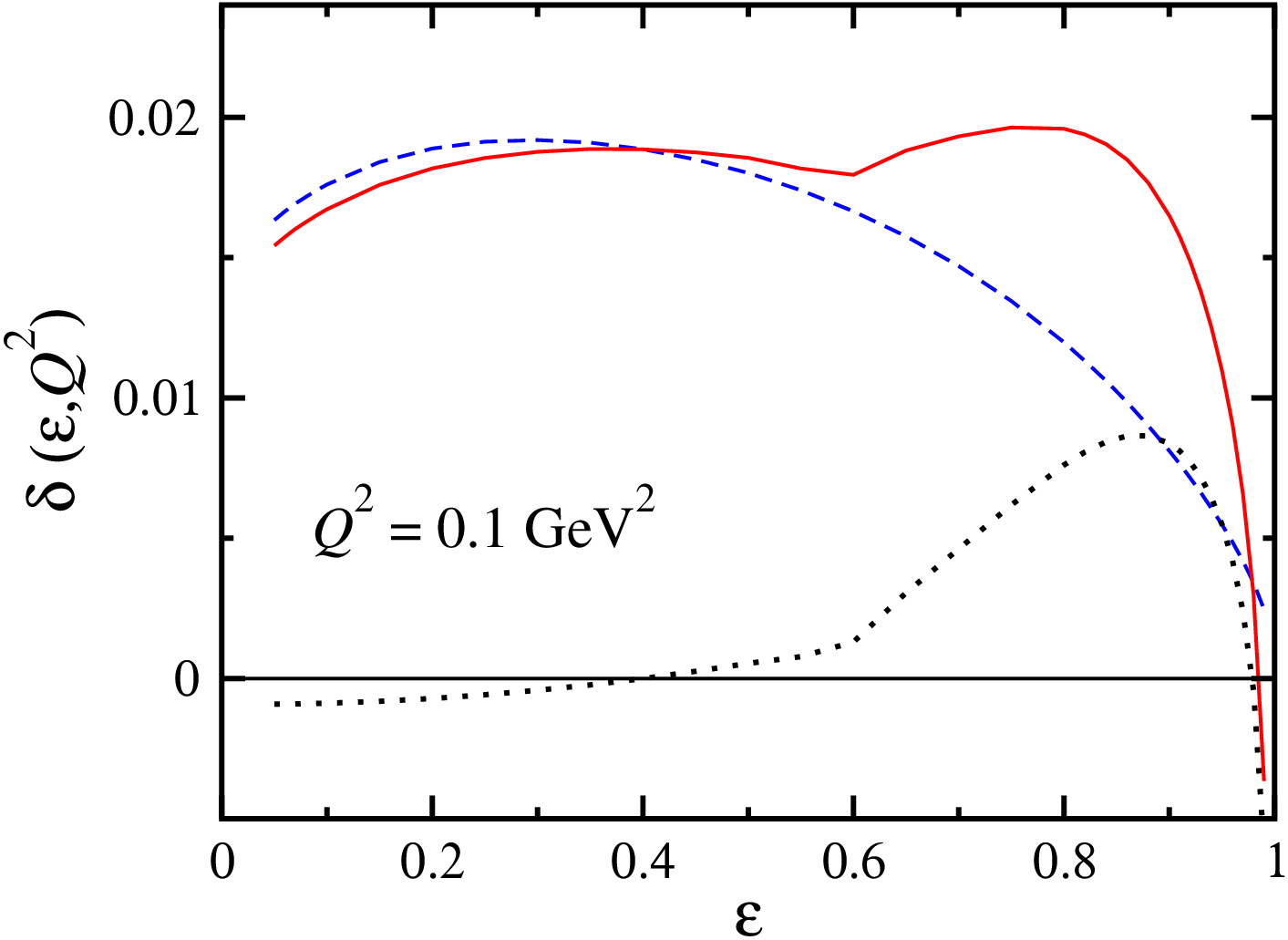}\vspace*{1.5cm}
\includegraphics[height=5.5cm]{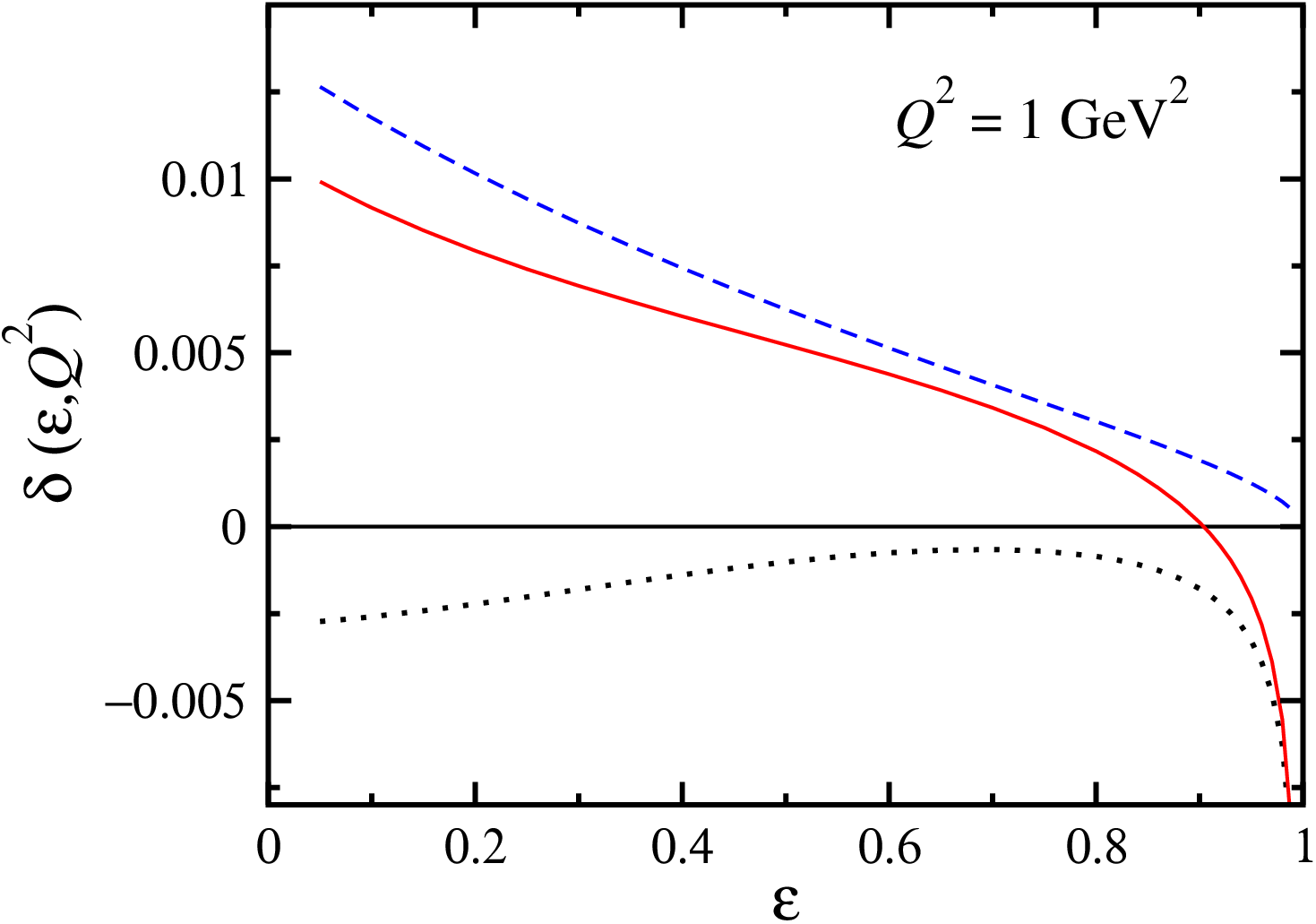}\hspace*{0.5cm}
\includegraphics[height=5.5cm]{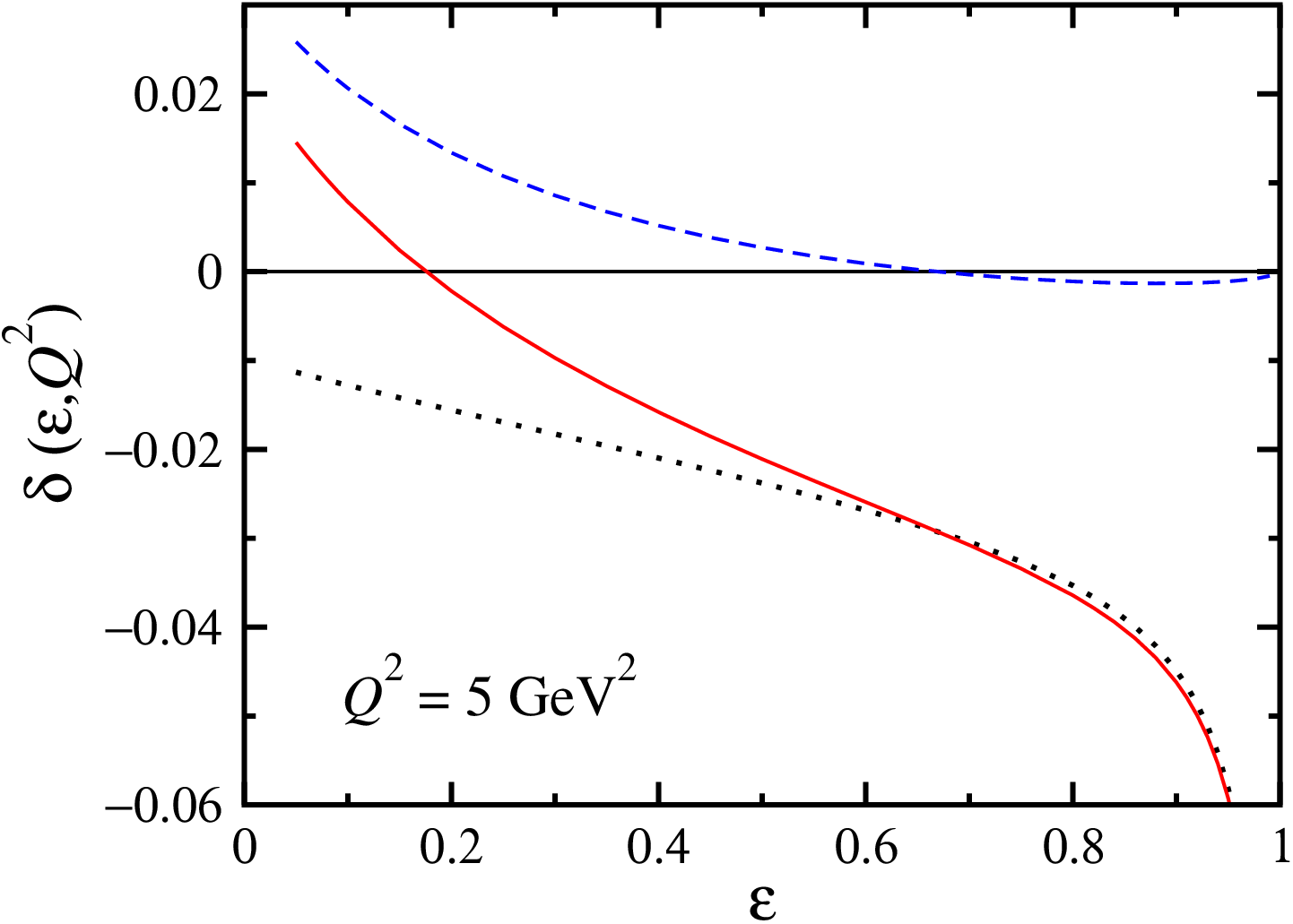}
\caption{TBE corrections $\delta_N(\eps,Q^2)$ for the nucleon (dashed)
	and $\delta_\Delta(\eps,Q^2)$ for the $\Delta(1232)$ (dotted)
	intermediate states, and the sum (solid),
	for $Q^2=0.01$, 0.1, 1 and 5~GeV$^2$.}
\label{fig:delND}
\end{figure}

A comparison of the total TBE corrections with nucleon and
$\Delta(1232)$ intermediate states, together with their sum,
is presented in Fig.~\ref{fig:delND} for $Q^2 = 0.01$, 0.1, 1
and 5~GeV$^2$.
As observed in the previous section, at small $\eps$
($\eps \lesssim 0.6$) the TBE correction at $Q^2 \lesssim 1$~GeV$^2$
is dominated by the nucleon elastic contribution.
At larger $\eps$ the $\Delta$ plays an increasingly important role,
and generally exceeds the nucleon piece at $\eps \gtsim 0.9$.
At higher $Q^2$, the magnitude of the $\Delta$ contribution is larger
than that of the nucleon for most $\eps$ values, although as remarked
above, the reliability of a purely resonant description of TBE is
less clear at momentum transfers above $Q^2 \sim 5$~GeV$^2$.

\begin{figure}[htb]
\includegraphics[height=7.5cm]{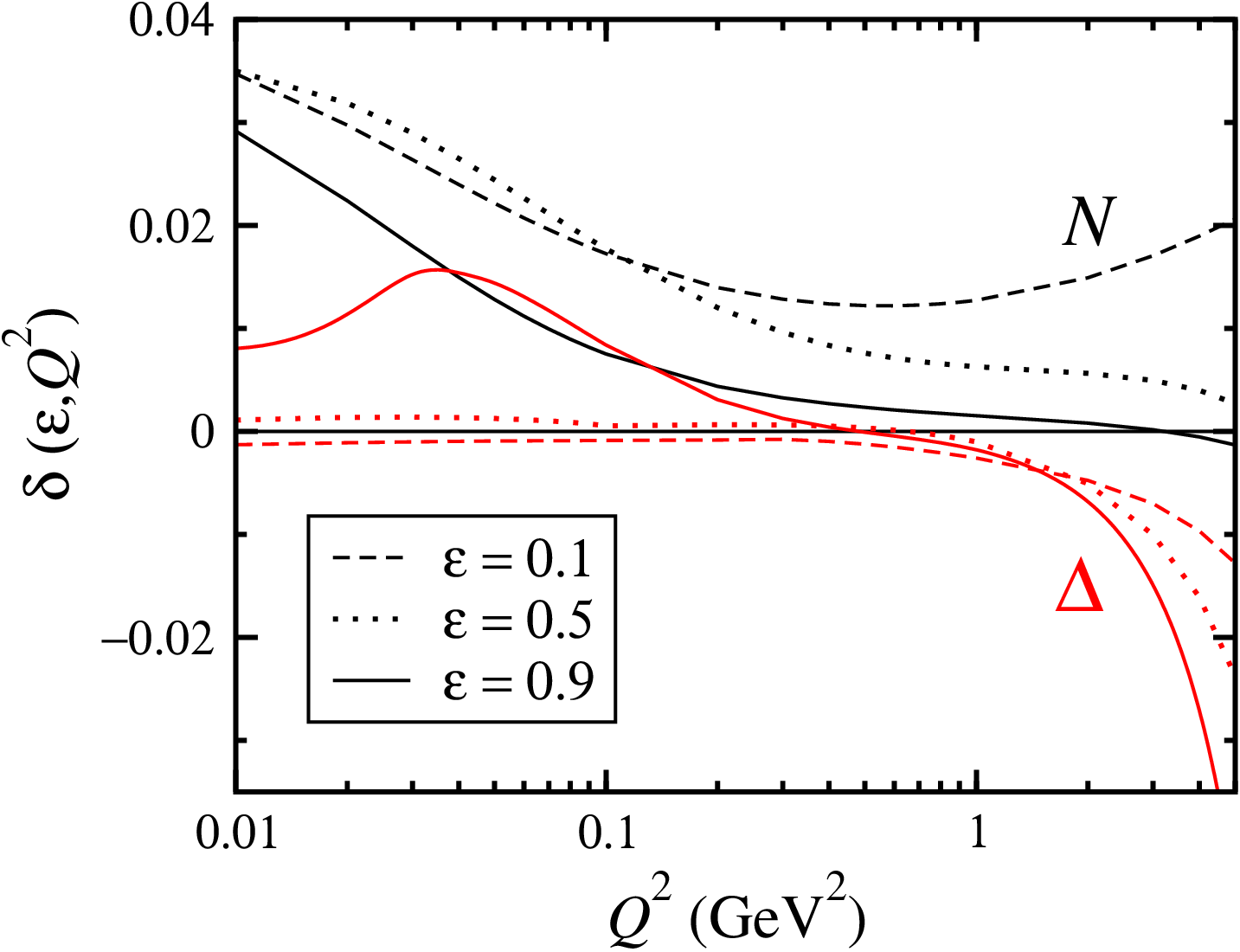}
\caption{Total TBE corrections $\delta_N$ (upper three curves)
	and $\delta_\Delta$ (lower three curves) versus $Q^2$ for
	fixed $\eps$ values, $\eps = 0.1$ (dashed), 0.5 (dotted)
	and 0.9 (solid).}
\label{fig:Qdep}
\end{figure}

The $Q^2$ dependence is more clearly illustrated in
Fig.~\ref{fig:Qdep}, where we show the nucleon and $\Delta$
corrections for fixed $\eps = 0.1$, 0.5 and 0.9.
At low $Q^2$ the nucleon correction $\delta_N$ increases as
$Q^2 \to 0$, but flattens out somewhat for larger $Q^2$.
The $\Delta$ correction $\delta_\Delta$, in contrast, is almost
$Q^2$ independent for $Q^2 \lesssim 1$~GeV$^2$, except at very high
$\eps$, but rapidly becomes large and negative at higher $Q^2$.

The results for $\delta_\Delta$ are different in shape and
magnitude from those reported by Nagata {\em et al.} \cite{YangD},
with the differences more pronounced at large $Q^2$.
As observed in Figs.~\ref{fig:delD_model} and \ref{fig:delD_g3},
the dependence on the input form factors and $N\Delta$ couplings
is unlikely to account for these differences.
We have checked the numerical calculations of the TBE amplitudes
using two independent computer codes, and find agreement between them.
It is not clear therefore what the origin of the differences may be.
Nevertheless, we do agree with the general finding in
Ref.~\cite{YangD} that the $\Delta$ plays an increasingly important
role at forward angles compared with the nucleon.

\begin{figure}[t]
\includegraphics[height=7.5cm]{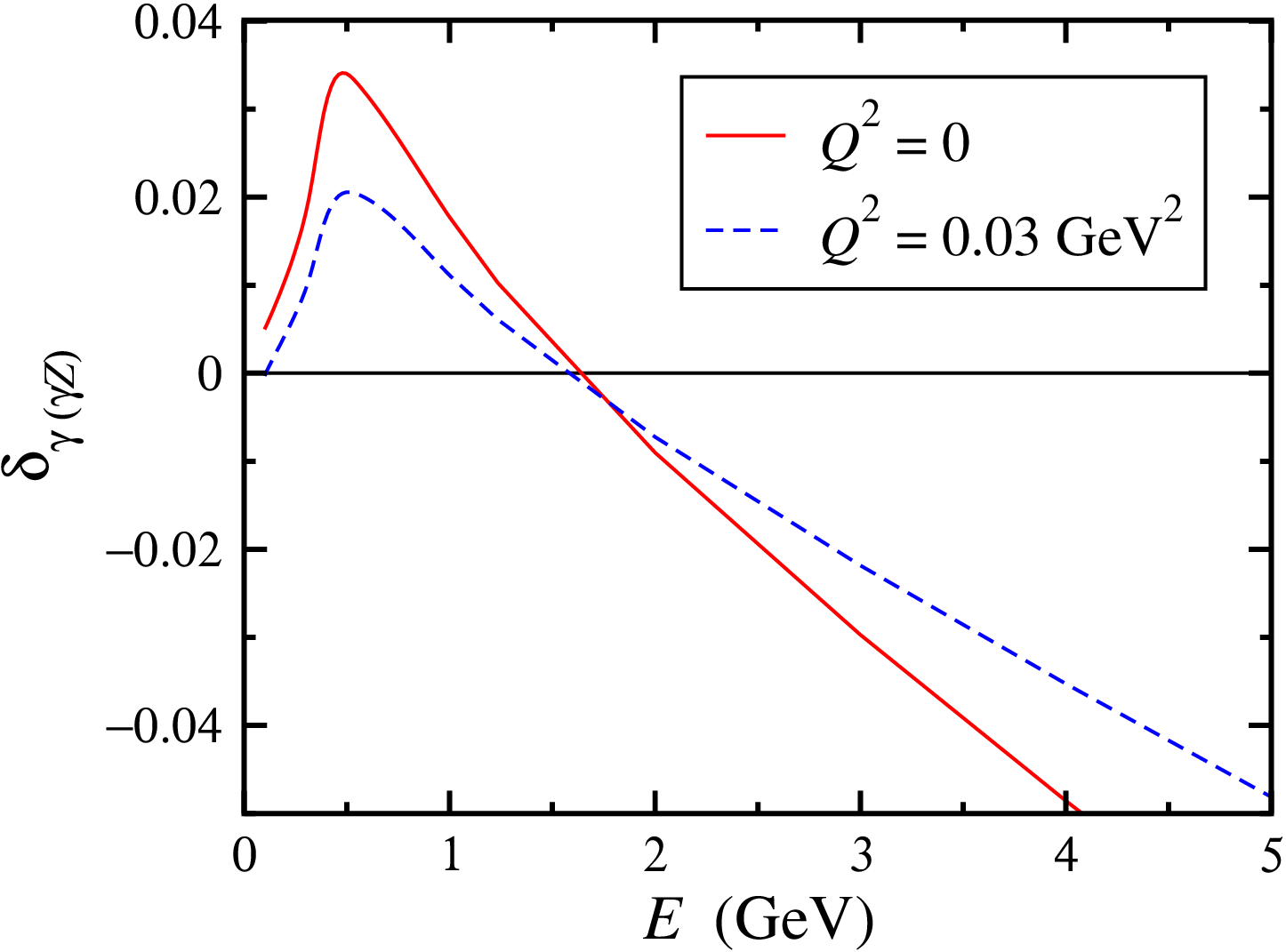}
\caption{TBE correction $\delta_{\gamma(\gamma Z)}$ arising from
	nucleon and $\Delta$ intermediate states as a function
	of the incident electron energy $E$, for $Q^2=0$ (solid)
	and $Q^2=0.03$~GeV$^2$ (dashed).}
\label{fig:del_E}
\end{figure}

While the $\Delta$ correction is relatively small for $Q^2$ between
around 0.01 and 3~GeV$^2$, at very low $Q^2$ there can be a sizable
enhancement of the $\gamma Z$ contribution at extremely forward angles,
$\eps \to 1$, corresponding to large incident electron energies.
This point was made recently in Ref.~\cite{GoHo}, which argued for a
large inelastic Regge contribution in the high energy limit.
In this region the TPE contribution is suppressed, and the Born term
is dominated by the proton weak charge, $Q_w$.
Hence the $\Delta$ contribution would be enhanced by a factor
$(1+Q_w)/Q_w \approx 14$.
In Fig.~\ref{fig:del_E} we show the sum of the nucleon and $\Delta$
contributions to $\delta_{\gamma(\gamma Z)}$ as a function of the
incident electron energy, for $Q^2=0$ and for the Qweak \cite{QWEAK}
value $Q^2=0.03$~GeV$^2$.
The $\Delta$ contribution rises linearly with energy up to 
$E \sim 0.5$~GeV, where it reaches $\approx 2-3\%$, after which
it decreases.
This is qualitatively similar to the resonance contributions found
in Ref.~\cite{GoHo}.

\begin{table}
\caption{TBE corrections for the nucleon ($\delta_N$) and $\Delta$
	($\delta_\Delta$) intermediate states, and their sum
	(in percent), at various experimental kinematics.
	Also shown are corrections from removing the existing hadronic
	$\delta_{\rm MS}^{\rm had}$ and total (hadronic + asymptotic)
	$\delta_{\rm MS}^{\rm tot}$ corrections at $Q^2=0$
	\cite{MS,PDG}.}
\begin{tabular}{ccc|ccc|cc}
$Q^2$~(GeV$^2$) & $\theta$ & Expt.&\ \ \ $\delta_N$\ \ \
				  & $\delta_\Delta$
				  & $\delta_{N+\Delta}$
				  &\ \ \ $\delta_{\rm MS}^{\rm had}$\ \ \
				  & $\delta_{\rm MS}^{\rm tot}$
							\\ \hline\hline
0.099	& 6.0$^\circ$	& \ \ \ HAPPEX \cite{HAPPEX04}\ \ \ 
				& 0.19	&$-1.20$  & $-1.01$
				& 0.45	& 2.42		\\
0.477	& 12.3$^\circ$  & \ \ \ HAPPEX \cite{HAPPEX04}\ \ \ 
				& 0.13	&$-0.44$  & $-0.31$
				& 0.16	& 0.86		\\ \hline
0.077	& 6.0$^\circ$ 	& \ \ \ HAPPEX \cite{HAPPEX07}\ \ \ 
				& 0.22 	&$-1.04$  & $-0.82$
				& 0.52	& 2.78		\\ \hline
0.1     & 144.0$^\circ$ & \ \ \ SAMPLE \cite{SAMPLE97}\ \ \
				& 1.63	&$-0.09$  & 1.54
				& 0.06	& 0.33		\\ \hline
0.108	& 35.37$^\circ$ & PVA4 \cite{PVA404}
				& 1.05	& 0.78    & 1.83
				& 0.37	& 1.98		\\
0.23	& 35.31$^\circ$ & PVA4 \cite{PVA404}
				& 0.62	& 0.34    & 0.96
				& 0.23	& 1.22		\\ \hline
0.224	& 145.0$^\circ$ & PVA4 \cite{A4back}
				& 1.33	&$-0.07$  & 1.27
				& 0.06	& 0.30		\\ \hline
0.122	& 6.68$^\circ$  & G0 \cite{G0}
				& 0.18	&$-1.06$  & $-0.88$
				& 0.40	& 2.13		\\
0.128	& 6.84$^\circ$	& G0 \cite{G0}
				& 0.18	&$-1.03$  & $-0.85$
				& 0.39	& 2.07		\\
0.136	& 7.06$^\circ$  & G0 \cite{G0}
				& 0.18	&$-0.99$  & $-0.81$
				& 0.37	& 1.99		\\
0.144	& 7.27$^\circ$  & G0 \cite{G0}
				& 0.17	&$-0.96$  & $-0.79$
				& 0.36	& 1.92		\\
0.153	& 7.5$^\circ$   & G0 \cite{G0}
				& 0.17	&$-0.92$  & $-0.75$
				& 0.35	& 1.85		\\
0.164	& 7.77$^\circ$  & G0 \cite{G0}
				& 0.17	&$-0.88$  & $-0.71$
				& 0.33	& 1.77		\\
0.177	& 8.09$^\circ$  & G0 \cite{G0}
				& 0.16	&$-0.83$  & $-0.67$
				& 0.32	& 1.69		\\
0.192	& 8.43$^\circ$  & G0 \cite{G0}
				& 0.16	&$-0.79$  & $-0.63$
				& 0.30	& 1.60		\\
0.21	& 8.84$^\circ$  & G0 \cite{G0}
				& 0.16	&$-0.73$  & $-0.57$
				& 0.28	& 1.51		\\
0.232	& 9.31$^\circ$  & G0 \cite{G0}
				& 0.16	&$-0.68$  & $-0.52$
				& 0.26	& 1.41		\\
0.262	& 9.92$^\circ$  & G0 \cite{G0}
				& 0.15	&$-0.62$  & $-0.47$
				& 0.24	& 1.30		\\
0.299	& 10.63$^\circ$ & G0 \cite{G0}
				& 0.15	&$-0.55$  & $-0.40$
				& 0.22	& 1.19		\\
0.344	& 11.46$^\circ$ & G0 \cite{G0}
				& 0.15	&$-0.48$  & $-0.33$
				& 0.20	& 1.07		\\
0.41	& 12.59$^\circ$ & G0 \cite{G0}
				& 0.15	&$-0.41$  & $-0.26$
				& 0.18	& 0.95		\\
0.511	& 14.2$^\circ$  & G0 \cite{G0}
				& 0.15	&$-0.32$  & $-0.17$
				& 0.15	& 0.81		\\
0.631	& 15.98$^\circ$ & G0 \cite{G0}
				& 0.15	&$-0.26$  & $-0.11$
				& 0.13	& 0.70		\\
0.788	& 18.16$^\circ$ & G0 \cite{G0}
				& 0.16	&$-0.23$  & $-0.07$
				& 0.11	& 0.60		\\
0.997	& 20.9$^\circ$  & G0 \cite{G0}
				& 0.17	&$-0.22$  & $-0.05$
				& 0.10	& 0.51		\\ \hline
0.23	& 110.0$^\circ$ & G0 \cite{G0back}
				& 1.37	&$-0.10$  & 1.27
				& 0.09	& 0.47		\\
0.62	& 110.0$^\circ$ & G0 \cite{G0back}
				& 1.10	&$-0.15$  & 0.95
				& 0.07	& 0.35		\\ \hline
0.03	& 8.0$^\circ$   & Qweak \cite{QWEAK}
				& 0.57	&$-0.45$  & 0.13
				& 0.80	& 4.25		\\
\end{tabular}
\label{tab:exp}
\end{table}

The corrections to the $A_{\rm PV}$ asymmetry at kinematics
corresponding to past and planned experiments
\cite{HAPPEX04,G0,HAPPEX07,G0back,SAMPLE97,PVA404,A4back,QWEAK}
are listed
in Table~\ref{tab:exp}, where the nucleon ($\delta_N$) and $\Delta$
($\delta_\Delta$) contributions, together with their sum, are shown
(in percent \%) for various $Q^2$ and laboratory scattering angles
$\theta$.
In the numerical calculations the empirical proton \cite{AMT} and
neutron \cite{Bosted} electromagnetic form factors are used, with
dipole parameterizations for the axial form factors, as discussed
in Sec.~\ref{sec:TBE}.

For the forward angle HAPPEX \cite{HAPPEX04} and G0 \cite{G0}
measurements, the nucleon correction $\delta_N$ is in the
vicinity of $\sim 0.1 - 0.2 \%$, but increases to $\sim 1.0 - 1.5 \%$
for the backward angle G0 \cite{G0back} and the earlier SAMPLE
\cite{SAMPLE97} measurements.
In contrast, at forward kinematics the $\Delta$ contribution
$\delta_\Delta$ is negative and of order $-0.5\%$ to $-1\%$,
but is almost negligible ($\sim -0.1\%$) at backward angles.

\pagestyle{empty}

When combined, the results reveal a nontrivial interplay between the
total nucleon and $\Delta$ contributions, with the nucleon dominating
the backward angle corrections, and the $\Delta$ contribution driving
the forward angle kinematics, where it is rapidly varying with both
$\eps$ and $Q^2$.
Consequently, at the intermediate angles $\theta \approx 35^\circ$ of
the PVA4 experiment both the $N$ and $\Delta$ corrections are positive,
and combine to give a net $\sim 1-2\%$ effect.
For the planned Qweak experiment \cite{QWEAK} at very low $Q^2$
($= 0.03~$GeV$^2$) and $\theta = 8^\circ$, on the other hand, the
positive nucleon and negative $\Delta$ contributions mostly cancel,
leaving a much smaller overall correction of $\sim 0.1\%$.

Before correcting the experimental asymmetries for the above TBE
effects, one should note that the standard data analyses do already
include an estimate of TBE effects \cite{MS,PDG}.
These are usually taken from the classic analysis of Marciano \& Sirlin
\cite{Marciano,MS} who computed the $\gamma(Z\gamma)$ contributions at
$Q^2=0$.
Recent explicit calculations \cite{Yang,TM}, however, have found a
strong $Q^2$ dependence at small values of $Q^2$, which could
significantly impact the extrapolation of the $Q^2=0$ results to the
experimental kinematics.
In order to implement the full $Q^2$ dependence of the TBE corrections,
and avoid double counting of the effects in the data analyses, one must
remove the $Q^2=0$ TBE corrections, which are usually parameterized in
terms of $\rho$ and $\kappa$ \cite{MS,PDG}, before adding the
corrections computed here.

In Ref.~\cite{MS} the loop integration in the box diagram is broken up
into a ``hadronic'', low-mass part and an ``asymptotic'', high-mass
contribution given by
\be
K^{\rm asy}\
=\ M_Z^2 \int_{\mu^2}^\infty dk^2 {1 \over k^2 (k^2 + M_Z^2)}\
=\ \log{M_Z^2 \over \mu^2}\
+\ {\cal O}\left( {\mu^2 \over M_Z^2} \right)\ ,
\ee
where $\mu$ is the cut-off mass which defines the mass separation,
typically of the order of 1~GeV.
For $\mu \approx 0.5-1$~GeV, $K^{\rm asy}$ is in the range
$\approx 8-10$.
The hadronic part is computed in Ref.~\cite{MS} at $Q^2=0$ using
dipole form factors.

\pagestyle{plain}

To assess the effect of the new TBE contribution, we display in
Table~\ref{tab:exp} the corrections $\delta_{\rm MS}$ (in percent)
defined as
\begin{eqnarray}
\delta_{\rm MS} &=& 
{ A_V(\rho,\kappa)
  - A_V(\rho-\Delta\rho_{\rm MS},\kappa-\Delta\kappa_{\rm MS})
\over A_V(\rho,\kappa) }\ ,
\end{eqnarray}
where the numerical values for the $\Delta\rho_{\rm MS}$ and
$\Delta\kappa_{\rm MS}$ corrections (for $\mu = 1$~GeV) are
\begin{subequations}
\begin{eqnarray}
\left(	\Delta\rho_{\rm MS}^{\rm had},
	\Delta\kappa_{\rm MS}^{\rm had}
\right)
&=& (-0.07\%, -0.10\%)\ ,	\\
\left(
	\Delta\rho_{\rm MS}^{\rm tot},
	\Delta\kappa_{\rm MS}^{\rm tot}
\right)
&=& (-0.37\%, -0.53\%)\ ,
\end{eqnarray}
\end{subequations}
for the hadronic only and total (hadronic + asymptotic) contributions,
respectively.
The latter were subtracted in the analyses of Refs.~\cite{Yang,YangD},
whereas we believe that {\em only} the $Q^2=0$ {\em hadronic}
component should be removed when adding the new TBE corrections.
Numerically the hadronic contribution is much smaller than the
asymptotic, with the total $\delta_{\rm MS}^{\rm tot}$ being around
$1-3\%$ for forward kinematics, and over 4\% for the proposed Qweak
experiment \cite{QWEAK}.
The hadronic correction $\delta_{\rm MS}^{\rm had}$ is also largest
at forward angles, but is typically $0.1-0.4\%$ for most of the
experiments, and ranging up to 0.8\% for the Qweak kinematics.

The impact of these differences on the strange form factors is
difficult to gauge without performing a full reanalysis of the data,
since in general different electroweak parameters and form factors
are used in the various experiments
\cite{HAPPEX04,G0,HAPPEX07,G0back,SAMPLE97,PVA404,A4back}.
Following Zhou {\em et al.} \cite{Yang}, an estimate of the induced
difference between the strange asymmetry extracted using the different
form factors was made in Ref.~\cite{TM}.
Differences of the order of 15\% were found between the empirical and
monopole form factors (as used in Ref.~\cite{Yang}) for the HAPPEX
kinematics \cite{HAPPEX04,HAPPEX07}, around 20\% for the G0 datum
\cite{G0} in Table~I, and over 30\% for the PVA4 kinematics
\cite{PVA404}.
One should caution, however, that these values are indicative only,
and a more detailed reanalysis of the strange form factor data
including TBE effects is currently in progress \cite{GSrean}.

\section{Conclusion}
\label{sec:conc}

In this paper we have presented a comprehensive analysis of two-boson
($\gamma$ and $Z$) exchange corrections in parity-violating
electron--proton elastic scattering, paying particular attention
to the effects arising from the substructure of the nucleon.
Working within a hadronic framework, we have computed contributions
from box (and crossed box) diagrams in which the intermediate states
are described by nucleons and $\Delta$ baryons.

The $\Delta$ contribution is found to be much smaller than the nucleon
at small $\eps$, but becomes dominant at forward scattering angles.
The dependence of the corrections on the input hadronic form factors
is small for $Q^2 \lesssim 1$~GeV$^2$, but becomes appreciable at
higher $Q^2$ ($Q^2 \gtsim 5$~GeV$^2$), indicating the approximate
limit beyond which the hadronic calculations may no longer be reliable.

As well as studying their detailed $\eps$ and $Q^2$ dependence, we have
evaluated the nucleon and $\Delta$ TBE corrections relevant for recent
and planned parity-violating experiments
\cite{HAPPEX04,G0,HAPPEX07,G0back,SAMPLE97,PVA404,A4back,QWEAK}, finding
a nontrivial interplay between the $N$ and $\Delta$ contributions.
The total corrections at low $Q^2$ range from $\sim -1\%$ for forward
angles to $\sim 1-2\%$ at backward kinematics.
For the planned Qweak experiment \cite{QWEAK} we find a large
cancellation between the (positive) $\delta_N$ and (negative)
$\delta_\Delta$ corrections, resulting in a modest, $\sim 0.1\%$
effect overall.

Our results for the $\Delta$ differ significantly from those in
the recent analysis of Ref.~\cite{YangD}, with the correction
$\delta_\Delta$ differing both in sign and magnitude.
We have explored the possible origin of these differences by
studying the dependence of the corrections on the input nucleon
and $N \Delta$ transition form factors, but find the effects to be
much smaller than that needed to explain the discrepancy.
We also highlight the need for a careful treatment of the subtraction
of the standard Marciano-Sirlin $\gamma Z$ correction at $Q^2=0$
before adding the new contributions.
The results computed here can be used in future data analyses
to more reliably extract strange electromagnetic form factors
\cite{Young,GSrean} or standard model electroweak parameters
\cite{YoungSM}.

\begin{acknowledgments}

We are grateful to J.~Arrington, F.~Benmokhtar, O.~Lalakulich, 
V.~Pascalutsa and E.~Paschos for helpful discussions and 
communications.
W.~M. is supported by the DOE contract No. DE-AC05-06OR23177, under
which Jefferson Science Associates, LLC operates Jefferson Lab.

\end{acknowledgments}

\appendix
\section{Relations to Other $N\Delta$ Transition Form Factors}

In the literature other notations exist for the $N\Delta$ transition
form factors.
In this appendix we relate the form factors defined in this analysis
with those used elsewhere.

In Ref.~\cite{Caia04} (see also Refs.~\cite{PT04,JS})
the electromagnetic $\gamma N\Delta$ vertex is defined as
\begin{eqnarray}
\Gamma_{\gamma\Delta \to N}^{\mu\alpha}(p,q)
&=& {3 (M+M_\Delta) \over 2 M [(M+M_\Delta)^2 + Q^2]}\sqrt{2\over 3}\
\Big\{
   \bar{g}_M(Q^2)\
   \varepsilon^{\mu\alpha\nu\beta} p_\nu q_\beta\	\nonumber\\
& &
-\ \bar{g}_E(Q^2)
   \left[ p^\mu q^\alpha - g^{\mu\alpha} p\cdot q
   \right] i \gamma_5\					\nonumber\\
& &
-\ {\bar{g}_C(Q^2) \over M}
   \left[ q^2 \left( p^\mu \gamma^\alpha - g^{\mu\alpha} \fslash{p}
          \right)
        - q^\mu \left( q^\alpha \fslash{p} - \gamma^\alpha p\cdot q
        \right)
   \right] i \gamma_5
\Big\}\ .
\label{eq:Caia}
\end{eqnarray}
To relate this form to that in Eq.~(\ref{eq:gDN}), we note for
the $\bar{g}_M$ term the identity
\be
\epsilon^{\mu\nu\alpha\beta} \gamma_5 u_{\beta}(p)\
=\ \sigma^{\mu\nu} u^{\alpha}(p)\
-\ \sigma^{\mu\alpha} u^{\nu}(p)\
+\ \sigma^{\nu\alpha} u^{\mu}(p)\ ,
\label{eq:chis1}
\ee
where $\sigma^{\mu\nu}=\frac{i}{2}\,[\gamma^\mu,\,\gamma^\nu]$,
and $u^\alpha(p)$ is the Rarita-Schwinger spinor-vector for
the spin-3/2 $\Delta$ field.
Contracting with $p_\nu$ and $q_\alpha$ and making use of the
constraint relations
\be
p_\mu u^\mu(p) = 0\ ,\ \ \ \gamma_\mu u^\mu(p) = 0\ ,
\ee
one finds that the couplings are related by
\begin{subequations}
\begin{eqnarray}
g_{M,E}
&=& {-3 M_{\Delta}^2 \over M (M+M_\Delta)}\ \bar{g}_{M,E}\ ,    \\
g_C
&=& {-3 M_{\Delta}^3 \over M^2 (M+M_\Delta)}\ \bar{g}_C\ .
\end{eqnarray}
\end{subequations}

For the axial current, a vertex that one often encounters in the
literature is \cite{Paschos} (see also \cite{PaschosBook,Giessen})
\begin{eqnarray}
-i\ \Gamma_{ZN \to \Delta}^{\alpha\mu}(p,q)
&=& \frac{C_3^A(Q^2)}{M}
    \left( g^{\alpha\mu} \fslash{q} - q^\alpha \gamma^\mu \right)\
 +\ \frac{C_4^A(Q^2)}{M^2}
    \left( g^{\alpha\mu} p \cdot q - q^\alpha p^\mu \right)\
					\nonumber\\
& &
 +\ C_5^A(Q^2)\ g^{\alpha\mu}
 +\ \frac{C_6^A(Q^2)}{M^2}\ q^\alpha q^\mu\ ,
\label{eq:Paschos}
\end{eqnarray}
for an outgoing $\Delta$ with momentum $p$ and an incoming
$Z$ boson with momentum $q$.
Comparing with the expression in Eq.~(\ref{eq:ZDN_A}), and using the
Dirac equation, one finds the following relations for the form factors:
\begin{subequations}
\begin{eqnarray}
\frac{C_3^A}{M}
&=& \frac{1}{2 M_\Delta}\, g_1^A\ ,             \\
\frac{C_4^A}{M^2}
&=& \frac{1}{2 M_\Delta^2}\, \left( g_2^A - 2 g_1^A \right)\ ,  \\
C_5^A
&=& \frac{q^2}{2 M_\Delta^2}\, g_3^A\ ,           \\
\frac{C_6^A}{M^2}
&=& -\frac{1}{2 M_\Delta^2}\, g_3^A\ .
\end{eqnarray}
\end{subequations}
The form factors $C_5^A$ and $C_6^A$ are related by PCAC,
$C_6^A \to C_5^A\ M^2/Q^2$ in the chiral limit, with
$C_5^A(0) = f_\pi g_{\pi N \Delta}/\sqrt{3} = 1.2$.
The fit in Ref.~\cite{Paschos} to the neutrino $\Delta$-production
data gives $C_3^A = 0$ and $C_4^A = -C_5^A/4$, leaving a single unique
form factor, which is taken to be $C_5^A$.
One may therefore identify the axial couplings in
Eq.~(\ref{eq:ZDN_A}) as
\begin{subequations}
\begin{eqnarray}
g_1^A(Q^2) &=& 0\ ,                  			\\
g_2^A(Q^2) &=& -{M_\Delta^2 \over 2 M^2}\ C_5^A(Q^2)\ ,	\\
g_3^A(Q^2) &=& {2 M_\Delta^2 \over  q^2}\ C_5^A(Q^2)\ .
\end{eqnarray}
\end{subequations}
To compute the $g_3^A$ contribution, we include the $1/q^2$
factor in the form factor, and use the relation
$$
\frac{1}{q^2}\,\frac{1}{q^2-\Lambda^2}
= {1 \over \Lambda^2}
  \left( -\frac{1}{q^2} + \frac{1}{q^2-\Lambda^2} \right)\ .
$$


\end{document}